\documentclass[
reqno,12pt,oneside]{amsart}
\usepackage{amsmath,amssymb,amsfonts}
\usepackage{tikz-cd}
\usepackage{amsbsy}
\usepackage{amscd}
\usetikzlibrary{matrix,arrows,decorations.pathmorphing}
\usetikzlibrary{quantikz2}
\usepackage{graphicx}
\usepackage{float}
\usepackage{enumitem}
\usepackage{setspace}
\usepackage[margin=1.0in]{geometry}
\usepackage[all]{xy}

\usepackage{latexsym}

\usepackage{color}

\usepackage
{hyperref}
\hypersetup{%
  colorlinks,
  bookmarksopen,
  bookmarksnumbered,
  citecolor=blue,
  linkcolor=black,
  pdfstartview=FitH,
}

\newtheorem{theorem}{Theorem}[section]
\newtheorem{lemma}[theorem]{Lemma}
\newtheorem{proposition}[theorem]{Proposition}

\theoremstyle{definition}
\newtheorem{example}[theorem]{Example}

\newtheorem{remark}[theorem]{Remark}

\DeclareMathOperator{\id}{id}

\newcommand{\LTRN}{L^2(\mathbb{R}^n)}

\newcommand{\ten}{\otimes}

\newcommand{\vphi}{\varphi}

\newcommand{\A}{\mathcal{A}}

\newcommand{\C}{\mathbb{C}}
\newcommand{\N}{\mathbb{N}}
\newcommand{\Z}{\mathbb{Z}}
\newcommand{\R}{\mathbb{R}}
\newcommand{\T}{\mathbb{T}}

\DeclareSymbolFont{lettersA}{U}{txmia}{m}{it}
\DeclareMathSymbol{\W}{\mathord}{lettersA}{151}

\newcommand{\om}{\omega}
\newcommand{\Om}{\Omega}

\newcommand{\ad}{\mathrm{ad}}

\providecommand{\norm}[1]{\lVert#1\rVert}

\newcommand{\mc}[1]{\mathcal{#1}}

\newcommand{\la}{\langle}
\newcommand{\ra}{\rangle}
\newcommand{\rmv}[1]{}

\begin{document}

\title[]{The continuous variable Clifford hierarchy}

\author{Thomas Admiral}
\email{thomasadmiral@cmail.carleton.ca}
\address{School of Mathematics and Statistics, Carleton University, Ottawa, ON, Canada K1S 5B6}

\author{Jason Crann}
\email{jasoncrann@cunet.carleton.ca}
\address{School of Mathematics and Statistics, Carleton University, Ottawa, ON, Canada K1S 5B6}

\begin{abstract}
The Clifford hierarchy emerged through gate teleportation as a large class of gates that admit fault-tolerant implementations. While the original qubit  hierarchy extends naturally to qudit and continuous variable systems, its detailed mathematical structure has only been studied in the finite-dimensional setting. Motivated by its potential applications in quantum computing, in this paper, we initiate a structural investigation of the continuous variable Clifford hierarchy. Using different techniques, we establish analogous results to the qubit and qudit settings, including the characterization of diagonal gates in the hierarchy and the fact that all single-mode gates are semi-Clifford. In contrast to the qubit setting, we show that for any number of modes, all third-level gates in the hierarchy are semi-Clifford, thereby establishing the continuous variable analogue of a conjecture of Zeng-Chen-Chuang. Our results are illustrated with a number of examples, including non-semi-Clifford gates and the fact that the single-mode ($4^{th}$-order) Kerr evolution does not belong to any level of the hierarchy. 
\end{abstract}

\maketitle

\section{Introduction}

Gate teleportation is an important primitive in quantum computing, allowing for fault-tolerant implementations of a recursively defined class of gates known as the Clifford hierarchy \cite{GC}. Illuminating the mathematical structure of the hierarchy is an important problem within fault-tolerant quantum computing, and while the structure of the qubit and qudit Clifford hierarchies are not fully understood, many partial results have shed important insight, including the sufficiency of one-bit teleportation for the subclass of semi-Clifford gates \cite{dS,ZCC,ZLC}, the structure of third-level gates \cite{Aftab,BS,HRT,PRTC} and the characterizations of diagonal \cite{CGK} and 1- and 2-mode qudit gates \cite{CdS,dSL}.

Continuous variable (CV) gate teleportation, like its qudit counterpart, leads to fault-tolerant implementations of an analogous CV Clifford hierarchy \cite{BM}, and is relevant in photonic based modalities of quantum computing \cite{Meni,Metal,Wetal1}. By contrast, however, relatively little work has been done to understand the (higher-level) structure of the CV Clifford hierarchy. We aim to rectify this in this paper by initiating an in-depth study of its mathematical properties, highlighting important (non-)examples and practically motivated future directions.

Where the first level of the $n$-qudit Clifford hierarchy is defined as the Pauli group, the first level of the $n$-mode CV Clifford hierarchy, denoted $\mc{C}_1^n$, is the Weyl-Heisenberg group of displacememt operators, generated by linear combinations of the canonical observables $Q_1,...,Q_n, P_1,...,P_n$. For $k\ge2$, the $k^{th}$ level of the $n$-mode CV Clifford hierarchy, denoted $\mc{C}_k^n$, is the set of unitaries that send gates in $\mc{C}_1^n$ to gates in $\mc{C}_{k-1}^n$ under conjugation:
$$\mc{C}^n_k:=\{U\in\mc{U}(\LTRN)\mid U\mc{C}^n_1U^*\subseteq\mc{C}^n_{k-1}\}.$$
The second level, $\mc{C}_2^n$, coincides with the normalizer of $\mc{C}_1^n$, and is known as the ($n$-mode) Clifford group. Its elements can be implemented via linear optics and squeezing (see, e.g., \cite{FvL,Serafini} and the references therein).

While the higher-levels do not form groups, their subsets of diagonal gates do. In the CV setting, we define the diagonal hierarchy $\mc{D}^n_k$ as gates in $\mc{C}^n_k$ of the form $D=f(Q_1,...,Q_n)$, for a (measurable) function $f:\R^n\to\T$ (e.g., the single-mode cubic phase gate $e^{iQ^3}$). For $n$-qudit systems with $d$ a prime power, the diagonal gates in the hierarchy have been characterized through exponentials of certain polynomials \cite{CGK}. Our first result is an analogous characterization of the CV diagonal hierarchy (see Theorem \ref{t:diag}): 
\begin{theorem} For $n,k\in\N$, 
$$\mc{D}^n_k=\{e^{i p_k(Q_1,Q_2,...,Q_n)}\mid p_k \ \textnormal{multivariate polynomial of degree} \ \leq k\}.$$
\end{theorem}

Although directly analogous to \cite{CGK}, our proof techniques differ significantly, relying on measurable cocycles, continuous lifting criteria and distribution theory. 

In addition to their algebraic structure, gates in the diagonal hierarchy admit simpler implementations via ``one-bit'' teleportation \cite{dS,ZCC,ZLC}. Gates in the set $\mc{SC}_k^n:=\mc{C}^n_2\mc{D}^n_k\mc{C}^n_2$ admit a similar reduction in gate teleportation resources, and the resulting subclass is known as \textit{semi-Clifford gates} \cite{ZCC}. For $n=1,2$, all qubit and qudit ($d$ odd prime) hierarchy gates are semi-Clifford \cite{Aftab,CdS,dS,dSL,ZCC}, but when $k\ge4,n\ge3$, there exist gates that are not semi-Clifford \cite{ZCC}. It was conjectured that all third-level gates in any number of qubits are semi-Clifford \cite[Conjecture 1]{ZCC}, which was subsequently shown false by a counter-example of Gottesman-Mochon when $k=3$, $n=7$ (see \cite[\S1.1]{BS}).\footnote{All third-level qubit gates are so-called generalized semi-Clifford \cite{BS}, meaning there is an additional permutation gate in the decomposition. See \cite{dSL2} for a recent example of a 5-qubit gate that is not generalized semi-Clifford; thereby refuting \cite[Conjecture 2]{ZCC}.}

In Section \ref{s:semi-Clifford}, we show that all CV third-level gates are semi-Clifford in any number of modes (see Theorem \ref{t:SC}):

\begin{theorem}\label{t:12} For every $n\in\N$, 
$\mathcal{C}_3^n=\mc{SC}_3^n$.
\end{theorem}

This result is a positive solution to the continuous variable analogue of a conjecture of Zeng-Chen-Chuang \cite{ZCC}, marking an important distinction between the qubit and continuous variables settings. 

Since gates in the third level induce (smooth) Lie group homomorphisms $\mc{C}_1^n\to\mc{C}_2^n$, the proof of Theorem \ref{t:12} reduces to the structure of certain homomorphisms between their associated Lie algebras, which admits the following  classification (see Theorem \ref{t:classify}).

\begin{theorem} Let $n\in\N$. There is a one-to-one correspondence between $\mc{C}^n_3/\T$ and triples $(T,S,v)$ where 
\begin{itemize}
\item $T:\R^{2n}\times\R^{2n}\times\R^{2n}\to\R$ is a symmetric tri-linear map (i.e., symmetric 3-tensor) with isotropic support; 
\item $S\in Sp(2n,\R)$;
\item and $v\in\R^{2n}$.
\end{itemize}
\end{theorem} 

We remark that a similar classification was recently obtained independently in the qudit setting ($d$ an odd prime) \cite{Aftab}, which we became aware of when finalizing the manuscript. The results of \cite{Aftab} do not preclude possibility that every qudit gate in the third level is semi-Clifford. It would therefore be natural to examine whether our techniques shed insight into this question.

Portions of the literature on the CV Clifford hierarchy suggest that conjugation by a gate $U\in\mc{C}^n_k$ maps degree-$1$ polynomials in the canonical observables to degree-$(k-1)$ polynomials. Closer inspection of this suggested property leads to the technical hurdle of \textit{Schwartz preservation}: while every gate $U$ in the third level satisfies $U\mc{S}(\R^n)=\mc{S}(\R^n)$ (where $\mc{S}(\R^n)$ is the Schwartz space of rapidly decreasing smooth functions), it is not clear to the authors whether this property extends to $\mc{C}_k^n$, $k>3$. We therefore study the subhierarchy of gates, denoted $\mc{C}^{\infty,n}_k$, which preserve $\mc{S}(\R^n)$ in section \ref{s:smooth}. We show that any gate in the $k$-th level of this ``smooth'' hierarchy behaves as expected under conjugation: mapping degree-$1$ polynomials in the canonical observables to degree-$(k-1)$ polynomials. This infinitesimal property connects the smooth hierarchy to the automorphism group of the complex Weyl algebra $\mc{A}^n$ of all polynomials in the canonical observables. While the structure of $\mathrm{Aut}(\mc{A}^n)$ is not fully understood for general $n$, when $n=1$, it admits an amalgamated free product decomposition $\mathrm{Aut}(\mc{A}^1)=A\ast_{A\cap E} E$ with respect to the subgroups $A$ of affine symplectic automorphisms and $E$ of elementary triangular automorphisms (see, e.g., \cite{FM}). Utilizing this decomposition together with the geometry of the action on its associated Bass-Serre tree \cite{Serre}, we show that all single mode gates in the smooth hierarchy are semi-Clifford (see Theorem \ref{t:smoothSC}):

\begin{theorem} For all $k\in\N$, $\mc{C}^{\infty,1}_k=\mc{SC}^1_k$.
\end{theorem}

The qubit and qudit ($d$ prime) versions of this result were shown respectively in \cite{ZCC} and \cite{dSL}.

\section{Notation \& Preliminaries}

\subsection{The Weyl-Heisenberg Group} Throughout the paper we loosely follow the notational conventions of Folland \cite{Folland}, to which we refer the reader for additional background.

For $n\in\N$, let $H_n$ denote the $n$-mode \textit{Heisenberg group}, which is $\R^{2n+1}$ equipped with the group law
$$(a,b,t)\cdot(a',b',t')=(a+a',b+b',t+t'+\frac{1}{2}(ab'-a'b)), \ \ \ a,a',b,b'\in\R^n, \ t,t'\in\R,$$
where the dot product of two vectors in $\R^n$ is written by concatenation. The \textit{Schr\"{o}dinger representation} $\pi:H_n\to\mc{U}(L^2(\R^n))$ is given by
$$\pi(a,b,t)=e^{it}e^{\frac{i}{2}ab}M_bT_a,$$
where $M_b\psi(x)=e^{ibx}\psi(x)$ and $T_a\psi(x)=\psi(x+a)$ are the modulation and translation groups on $L^2(\R^n)$ (we take $\hbar\equiv 1$). The image $\mc{C}^n_1:=\pi(H_n)$, often called the \textit{Weyl-Heisenberg group}, is the continuous variable analogue of the Pauli group. It is well-known that $\pi$ is an irreducible, strongly continuous unitary representation of $H_n$. 

The associated Lie algbra $\mathfrak{h}_n$ is $\R^{2n+1}$ equipped with the Lie bracket
$$[(a,b,t),(a',b',t')]=(0,0,ab'-a'b), \ \ \ a,b\in\R^n.$$
Note that the final coordinate is the canonical symplectic form $$\Omega:\R^{2n}\times\R^{2n}\ni ((a,b),(a',b'))\mapsto ab'-a'b\in\R.$$
The space $\mc{H}^\infty(\pi)$ of smooth vectors for $\pi$ is known to be the Schwartz space
$$\mathcal{S}(\mathbb{R}^n):=\{\psi\in C^\infty(\mathbb{R}^n):\sup_{x\in\mathbb{R}^n}|(x^{\alpha}\partial_\beta\psi)(x)|<\infty \ \ \forall\alpha,\beta\in\mathbb{N}_0^n\}\subset L^2(\R^n)$$
(see, e.g., \cite[Theorem 2.2]{GT}). The family of semi-norms
$$p_{\alpha,\beta}(\psi):=\norm{x^{\alpha}\partial_\beta\psi}_{L^2(\R^n)}, \ \ \ \alpha,\beta\in\mathbb{N}_0^n,$$
induces a Fr\'{e}chet space structure on the subspace $\mc{S}(\R^n)\subset L^2(\R^n)$, under which the differentiated representation $d\pi$ maps $\mathfrak{h}_n$ to the space $\mc{L}(\mc{S}(\R^n))$ of continuous linear operators on $\mc{S}(\R^n)$ (see, e.g., \cite[\S10.1]{Schmudgen}). The image of the standard basis vectors $e_1,...,e_{n}\in\R^{n}$ yield the canonical observables in $n$-bosonic modes:
$$P_j\psi:=id\pi(e_j,0,0)\psi=-i\partial_j\psi, \ \ Q_j\psi:=id\pi(0,e_j,0)=x_j\psi, \ \ \ j=1,...,n, \ \psi\in\mc{S}(\R^n).$$
These observables satisfy the canonical commutation relations (CCR):
\begin{equation}\label{e:CCR}[Q_j,P_k]=i\delta_{j,k}1.\end{equation}
For a vector $(a,b)\in\R^{2n}$, the formal dot product with the tuple of operators $(P,Q):=(P_1,...,P_n,Q_1,...,Q_n)$ will be written as a concatenation:
$$(a,b)(P,Q):=aP+bQ=\sum_{j=1}^na_jP_j+b_jQ_j.$$
It follows that $\pi(a,b,t)=e^{it}e^{i(aP+bQ)}$, $(a,b,t)\in H_n$ \cite[pg. 22]{Folland}. Throughout the paper we occasionally adopt the shorthand notation $\pi(a,b):=\pi(a,b,0)$, $a,b\in\R^n$, and similarly for $d\pi$.

The next Proposition follows from known results (see, e.g., \cite[\S4]{Weil}), however we could not locate a detailed argument specific to this context. A proof is therefore included.

\begin{proposition}\label{p:SOT_1} For every $n\in\N$, $\mc{C}_1^n$ is strong operator closed in $\mc{U}(L^2(\R^n))$.
\end{proposition}

\begin{proof} Let $U\in\overline{\mc{C}_1^n}^{sot}\cap\mc{U}(L^2(\R^n))$. As $L^2(\R^n)$ is separable, the strong operator topology is metrizable on its unitary group, implying the existence of a sequence $((a_j,b_j,t_j))_j\in H_n$, such that $U=\lim_{j\to\infty} \pi(a_j,b_j,t_j)$. By strong continuity of the group operations in $\mc{U}(L^2(\R^n))$ and the canonical commutation relations, for every $(a,b)\in\R^{2n}$
\begin{align*}U\pi(a,b,0)U^*\pi(a,b,0)^*&=\lim_j \pi(a_j,b_j,t_j)\pi(a,b,0)\pi(a_j,b_j,t_j)^*\pi(a,b,0)^*\\
&=\lim_je^{i\Omega((a_j,b_j),(a,b))}1\in\T1.
\end{align*}
Thus, the map 
$$\chi:\R^{2n}\ni(a,b)\mapsto U\pi(a,b,0)U^*\pi(a,b,0)^*\in\T1$$
is a pointwise limit of (continuous) characters of $\R^{2n}$, thereby defining a (Haar) measurable homomorphism $\R^{2n}\to\T$. By Weil's automatic continuity theorem (see, e.g., the proof of \cite[Corollary 2.4]{Rosendal}), $\chi$ is continuous. Moreover, a simple application of the dominated convergence theorem shows that $e^{i\Omega((a_j,b_j),\cdot)}\to\chi$ in the weak* topology of $L^\infty(\R^{2n})=L^1(\R^{2n})^*$. Since the dual group $\widehat{\R^{2n}}$ of continuous characters $\R^{2n}\to\T$ is homeomorphic to $\R^{2n}$ under said topology \cite[Corollary 4.7]{Folland2}, there exists $(c,d)\in\R^{2n}$ such that $\Omega(a_j,b_j)\to(c,d)$. Then 
$$\lim_j e^{it_j}1=\lim_n U\pi(a_j,b_j,0)^*=U\pi(\Omega^{-1}(c,d),0)^*$$
exists in $\T$, so is of the form $e^{it}$, $t\in\R$. Hence, $U=\pi(\Omega^{-1}(c,d),t)\in\mc{C}_1^n$.  
\end{proof}

Since $\mc{C}^n_1=\pi(H_n)$ is a closed subgroup of the Polish group $(\mc{U}(L^2(\R^n)),sot)$, it is Polish (meaning separable and completely metrizable). Then $\pi:H_n\to\mc{C}^n_1$ is a continuous surjective homomorphism between Polish groups, and is therefore open \cite[Corollary 2.3.4]{Gao}. Hence, the first isomorphism theorem for topological groups applies to give the homeomorphism $\mc{C}^n_1\cong H_n/\mathrm{Ker}(\pi)\cong \R^{2n}\times\T$.

\subsection{The Clifford Group} The symplectic group $Sp(2n,\R)$ is the group of linear maps $S:\R^{2n}\to\R^{2n}$ which preserve the sympelectic form, in that $\Omega(Sv,Sw)=\Omega(v,w)$ for all $v,w\in\R^{2n}$. Any such $S$ is automatically invertible, and $Sp(2n,\R)$ is a closed subgroup of the special linear group $SL(2n,\R)$ of determinant-1 matrices. When $n=1$, $Sp(2,\R)=SL(2,\R)$.

Any $S\in Sp(2n,\R)$ induces an automorphism of $H_n$ which acts trivially on the center via $(a,b,t)\mapsto(S(a,b),t)$. The composition $\pi\circ S:H_n\to\mc{U}(L^2(\R^n))$ is an irreducible (strongly continuous) unitary representation of $H_n$ satisfying $\pi\circ S(0,0,t)=e^{it}$, $t\in\R$. By the Stone-von Neumann theorem \cite[Theorem 1.50]{Folland}, there is a unitary $\mu(S)\in\mc{U}(L^2(\R^n))$ for which $\pi\circ S=\mu(S)\pi(\cdot)\mu(S)^*$. By irreducibility of $\pi$, the unitary $\mu(S)$ is determined uniquely up to global phase. (In fact, the phases can be chosen up to factors of $\pm1$ \cite[pg 183]{Folland}.) The induced projective representation $\mu:Sp(2n,\R)\to\mc{U}(L^2(\R^n))$ is called the \textit{metaplectic representation} of $Sp(2n,\R)$. 

The Lie algebra $sp(2n,\R)$ of $Sp(2n,\R)$ consists of those $\A\in M_{2n}(\R)$ for which $\A\Omega$ is symmetric, and is isomorphic to the Lie algebra of real homogeneous quadratic polynomials on $\R^{2n}$ equipped with the Poisson bracket via $sp(2n,\R)\ni \A\mapsto \mc{P}_{\A}$, where
$$\mc{P}_{\A}(v)=-\frac{1}{2}v\A\Omega v, \ \ \ v\in\R^{2n}.$$
Writing $\A=\begin{bmatrix} A & B\\ C & -A^t\end{bmatrix}$ in block form, we have
$$\mc{P}_{\A}(a,b)=\frac{1}{2}aBa-aAb-\frac{1}{2}bCb, \ \ \ a,b\in\R^{n}.$$
The map $sp(2n,\R)\ni A\mapsto iP_{\A}(P,Q)\in\mc{L}(\mc{S}(\R^n))$ is a faithful representation by continuous skew-symmetric operators on $\mc{S}(\R^n)$, where 
$$P_{\A}(P,Q):=\frac{1}{2}\sum_{j,k=1}^nB_{j,k}P_jP_k-\frac{1}{2}\sum_{j,k=1}^nA_{j,k}(P_jQ_l+Q_lP_j)-\frac{1}{2}\sum_{j,k=1}^nC_{j,k}Q_jQ_k.$$
The Schwartz space $\mc{S}(\R^n)\subseteq\mc{H}^\infty(\mu)$ and for each $\A\in sp(2n,\R)$ we have 
\begin{equation}\label{e:quadratic}d\mu(\mc{A})\psi=\frac{d}{dt}\mu(e^{t\A})|_{t=0}=iP_{\A}(P,Q)\psi, \ \ \ \psi\in\mc{S}(\R^n),\end{equation}
where the sign of $\mu(e^{t\A})$ is determined by continuity in $t$ and $\mu(e^{0\A})=1$ \cite[Theorem 4.45]{Folland}.

Defining the semi-direct product $HSp(2n,\R):=Sp(2n,\R)\ltimes H_n$ via the canonical action $Sp(2n,\R)\curvearrowright H_n$, it follows that 
$$\om:HSp(2n,\R)\ni (S,(a,b,t))\mapsto\mu(S)\pi(a,b,t)\in\mc{U}(L^2(\R^n))$$
lifts to a strongly continuous representation $\tilde{\omega}$ of the universal cover  $\widetilde{HSp(2n,\R)}$, known as the extended metaplectic representation \cite[Proposition 4.58]{Folland}. The image 
$$\mc{C}^n_2:=\tilde{\om}(\widetilde{HSp(2n,\R)})=\om(HSp(2n,\R))$$ 
is called the \textit{Clifford group}, and coincides with $\{U\in\mc{U}(L^2(\R^n))\mid U\mc{C}^n_1U^*=\mc{C}^n_1\}$, the normalizer of the Weyl-Heisenberg group (see \cite[pg 196-197]{Folland}, \cite[Theorem 1]{BSBN}). Since $\mc{C}^n_1$ is strong operator closed (Proposition \ref{p:SOT_1}), it follows that $\mc{C}^n_2$ is strong operator closed, and the same open mapping theorem argument from the previous section shows that $$\mc{C}^n_2\cong\widetilde{HSp(2n,\R)}/\mathrm{Ker}(\tilde{\om})$$  
as topological groups.

The Lie algebra $hsp(2n,\R)$ of $HSp(2n,\R)$ is isomorphic to $sp(2n,\R)\ltimes\mathfrak{h}_n$, and to the Lie algebra of polynomials of degree $\leq 2$ on $\R^{2n}$ with the Poisson bracket. The differential $d\om:hsp(2n,\R)\to\mc{L}(\mc{S}(\R^n))$ is $i$ times the canonical Weyl quantization of such polynomials (see \cite[\S2.1]{Folland}).

\section{The Clifford Hierarchy} 

As in the qudit setting, the continuous variable Clifford hierarchy \cite{BM} is defined recursively as follows. For a fixed number of modes $n\in\N$, the first level is the Weyl-Heisenberg group $\mc{C}^n_1=\pi(H_n)$, and for $k\geq 2$, the $k^{th}$-level of the $n$-mode Clifford hierarchy is  
$$\mc{C}^n_k:=\{U\in\mc{U}(\LTRN)\mid U\mc{C}^n_1U^*\subseteq\mc{C}^n_{k-1}\}.$$
Note that $\mc{C}^n_2$ defined in this way coincides with the Clifford group defined in the previous subsection.\footnote{If $U\in\mc{U}(L^2(\R^n))$ satisfies $U\mc{C}^n_1U^*\subseteq\mc{C}^n_1$, it induces a 
continuous (hence smooth) injective endomorphism on $\mc{C}^n_1\cong H_n/\mathrm{Ker}(\pi)$. Its associated differential is bijective by finite-dimensionality. Conjugation by $U$ is therefore bijective on a neighborhood of the identity in $H_n/\mathrm{Ker}(\pi)$, which, by connectivity, implies surjectivity on the whole group, so that $U\mc{C}^n_1U^*=\mc{C}^n_1$.}

We list some basic properties which follow from the definition:
\begin{itemize}
\item $\mc{C}^n_k\subseteq\mc{C}^n_{k+1}$ for all $n,k\in\N$.
\item $\T 1\subseteq\mc{C}^n_k$ for all $n,k\in\N$. 
\item $\mc{C}^n_k$ is a group for $k=1,2$ and $n\in\N$.
\item $\mc{C}^n_2\mc{C}^n_k\mc{C}^n_2\subseteq\mc{C}^n_k$ for all $k\geq 2$ and $n\in\N$. 
\end{itemize}
The final claim is easily verified by induction over $k$ for fixed $n\in\N$. The base case ($k=2$) is immediate. If the claim holds for $k$ then given $U\in\mc{C}^n_{k+1}$, $S,T\in\mc{C}^n_2$, and $W\in\mc{C}^n_1$, we have $SUTWT^*U^*S^*\in\mc{C}^n_{k}$. Indeed, $TWT^*\in\mc{C}^n_1$, so $UTWT^*U^*\in\mc{C}^n_{k}$ and therefore $SUTWT^*U^*S^*\in\mc{C}^n_{k}$ by the inductive hypothesis.

\begin{example} As exponentials of quadratic polynomials in the canonical observables, the following well-known classes of gates belong to the Clifford group $\mc{C}^n_2$:
\begin{itemize}
\item Phase space rotations $R_j(\theta)=e^{\frac{i\theta}{2}(Q_j^2+P_j^2)}$
\item Controlled phase gates $CQ(t)=e^{itQ_jQ_k}$
\end{itemize}
Similarly, the cubic phase gates $V(t)=e^{\frac{it}{3}Q_j^3}$ belong to the third level $\mc{C}^n_3$.
\end{example}

One might expect unitary groups $U_t=e^{itH(P,Q)}$ generated by degree-$k$ polynomials $H(P,Q)$ in the canonical observables to belong to (or at least be related to) the $k^{th}$-level of the CV Clifford hierarchy \cite{BM,vL}. Though we later show that this is not the case (see Example \ref{e:kerr}), the following Proposition provides a large class of gates for which it is true, including $H(P,Q)$ of the form $p_k(Q_1,...,Q_n)+cP$, for some polynomial $p_k$ and $c\in\R^{n}$.

\begin{proposition}\label{p:inclusion} Let $v_1,...,v_m\in\R^{2n}$ be linearly independent such that $\Omega(v_i,v_j)=0$ for all $i,j=1,...,m$ (i.e., $\mathrm{span}\{v_i\}$ is isotropic), and let $a,b\in\R^{n}$. Then for any $k\in\N$, and polynomial $p_k:\R^m\to\R$ degree $k$, $$H=p_k(v_1(P,Q),...,v_m(P,Q))+aP+bQ$$ is essentially self-adjoint on $\mc{S}(\R^n)$ and $U_t=e^{i t H}\in \mc{C}^n_k$ for all $t\in\R$. 
\end{proposition}

\begin{proof} By \cite[Lemma 2.17]{Folland}, it follows that $m\leq n$ and there exists $S\in Sp(2n,\R)$ such that $\mu(S)H\mu(S^{-1})=p_k(Q_1,...,Q_m)+S(a,b)(P,Q)$. Since $\mc{S}(\R^n)$ is an invariant subspace for $\mu$ \cite[Lemma 2.14]{Folland}, and each $\mc{C}_k^n$ is invariant under conjugation by $\mc{C}^n_2$, we may assume without loss of generality that all operators $H$ in the statement of the Proposition are of the form
$$H=p_k(Q_1,...,Q_n)+cP,$$
for some $c\in\R^n$ (by absorption of the $Q$-linear factor into $p_k$).

Let us proceed by induction on $k\in\N$. For $k=1$, $H$ is of the form $aP+bQ$ for some $a,b\in\R^{n}$, so that $U_t=e^{i t H}\in \mc{C}^n_1$ for all $t\in\R$. 

Suppose the claim holds for $k-1$. In this case, $iH=i(p_k(Q_1,...,Q_m)+cP)$
belongs to 
$$\mathfrak{g}_k:=\mathrm{span}_{\R}\{iP_j,iQ^\alpha, \mid j=1,...,n,\  \alpha\in\N^m_0, \ |\alpha|_1\leq k\},$$
a finite-dimensional Lie algebra of skew-symmetric operators on $\mc{S}(\R^n)$. Noting the non-zero commutation relations between its generators,
$$[iP_j,iQ^\alpha]=i\partial_jQ^\alpha=i\alpha_jQ^{\alpha-e_j},$$
it follows that $\mathfrak{g}_k$ is $(k+1)$-nilpotent in the sense that any $(k+1)$-order commutator is zero. The (negative of the) Nelson Laplacian (see \cite[\S9]{Nelson}) of $\mathfrak{g}_k$ is
$$-A_k:=\sum_{j=1}^n P_j^2+\sum_{|\alpha|_1\leq k} Q^{2\alpha}=-\Delta_n + V_k,$$
where $\Delta_n$ is the Laplacian of $\R^n$ and $V_k=\sum_{|\alpha|_1\leq k} Q^{2\alpha}$. Since $V_k\in L^2_{loc}(\R^n)$ (locally square integrable) and $V_k\geq 0$ pointwise, $-A_k$, and therefore $A_k$ is essentially self-adjoint on $C_c^\infty(\R^n)$ \cite[Theorem X.28]{RS2}. As $A_k$ is also symmetric on $\mc{S}(\R^n)$, it follows that $A_k$ is essentially self-adjoint on $\mc{S}(\R^n)$. Since, in addition, $\mathfrak{g}_k$ leaves $\mc{S}(\R^n)$ invariant, by \cite[Theorem 5]{Nelson} (and its proof), $iY$ is essentially self-adjoint on $\mc{S}(\R^n)$ for all $Y\in\mathfrak{g}_k$ and there is a (strongly continuous) unitary representation $\vphi:G_k\to\mc{U}(L^2(\R^n))$ of the unique simply connected Lie group $G_k$ with Lie algebra $\mathfrak{g}_k$ such that 
$$\vphi(\exp(tY))=e^{tY}, \ \ \ Y\in\mathfrak{g}_k, \ t\in\R.$$
By smoothness of the (finite-dimensional) adjoint representation $\mathrm{Ad}:G_k\to GL(\mathfrak{g}_k)$, and $(k+1)$-nilpotency of $\mathfrak{g}_k$, Taylor expansion yields 
$$\mathrm{Ad}_{\exp(tY)}(Z)=\sum_{l=0}^k\frac{1}{l!}\mathrm{ad}^l_{tY}(Z), \ \ \ Y,Z\in\mathfrak{g}_k, \ t\in\R.$$
Taking $Y=iH$ and $Z=i (aP+bQ)\in\mathfrak{g}_k$ for $a,b\in\R^n$, the canonical commutation relations imply 
\begin{equation}\label{e:dirder}[tY,Z]=it(-cb1+a\nabla p_k(Q_1,...,Q_n)),\end{equation}
from which one easily deduces that for each $1\leq l\leq k$, 
$$\ad_{tY}^l(Z)=\underbrace{[tY,[\cdots[tY}_{\textnormal{\textit{l} times}},Z]\cdots]$$ is $i$ times a polynomial of degree $k-l$ in $Q_1,..,Q_m$. Thus,
\begin{align*}\mathrm{Ad}_{\exp(i tH)}(i (aP+bQ))&=\sum_{l=0}^k\frac{1}{l!}\underbrace{[it H,[\cdots,[it H}_{\textnormal{\textit{l} times}}, i(aP+bQ)]\\
&=i((aP+bQ)+p_{k-1}(Q_1,...,Q_n))
\end{align*}
for some polynomial $p_{k-1}$ of degree $k-1$. Then
\begin{align*}U_t\pi(a,b,0)U_t^*&=e^{ itH}e^{i(aP+bQ)}e^{-it H}\\
&=\vphi(\exp(\mathrm{Ad}_{\exp(i tH)}(i(aP+bQ))))\\
&=e^{i((aP+bQ)+p_{k-1}(Q_1,...,Q_m))}.
\end{align*}
By the induction hypothesis, $e^{i((aP+bQ)+p_{k-1}(Q_1,...,Q_n))}\in\mc{C}^n_{k-1}$, and the proof is complete.
\end{proof}

The next proposition answers a natural topological question in this setting.

\begin{proposition} For every $k,n\in\N$, $\mc{C}_k^n$ is strong operator closed in $\mc{U}(L^2(\R^n))$.
\end{proposition}

\begin{proof} Towards induction, the base case ($k=1$) was shown in Proposition \ref{p:SOT_1}. 

Suppose the claim holds for $k-1$, and let $U\in\overline{\mc{C}_k^n}^{sot}\cap\mc{U}(L^2(\R^n))$. Then there is a sequence $(U_j)$ in $\mc{C}_k^n$ such that $U_j\to U$. By strong operator continuity of the group operations, for every $(a,b,t)\in H_n$,
$$U\pi(a,b,t)U^*=\lim_j U_j\pi(a,b,t)U_j^*\in\overline{\mc{C}_{k-1}^n}^{sot}\cap\mc{U}(L^2(\R^n))=\mc{C}_{k-1}^n.$$
Thus, $U\in\mc{C}_k^n$. 
\end{proof}

\section{The Diagonal Hierarchy}\label{s:diagonal}




The use of gate teleportation to implement a given unitary in the qubit hierarchy relies on the implementation of arbitrary Clifford group elements, computational basis measurements, and the injection of gate-dependent magic states. Diagonal gates $D$ admit a simpler, so-called one-bit teleportation circuit, wherein magic states are half the size \cite{ZCC,ZLC}.\footnote{In the qubit setting, this halving of resources is that between the diagonal image $D|+\ra$ of the Hadamard state $|+\ra$ and the local action $(1\ten D)|\psi_{00}\ra$ of $D$ on a maximally entangled Bell pair $|\psi_{00}\ra$.} One-bit teleportation circuits in the qudit setting were studied de Silva \cite{dS}.

Let $L^\infty(\R^n)$ denote the von Neumann algebra of Lebesgue measurable essentially bounded functions on $\R^n$. The mapping 
$M:L^\infty(\R^n)\ni f\mapsto M_f\in\mc{B}(L^2(\R^n)),$
where $M_f\psi(x)=f(x)\psi(x)$, is a weak*-weak* homeomorphic $*$-isomorphism onto its range \cite{KR}. The range $M(L^\infty(\R^n))$ is the continuous variable analogue of diagonal matrices, so for $n,k\in\N$, we define the diagonal Clifford hierarchy by $\mc{D}^n_k:=M(L^\infty(\R^n))\cap \mc{C}^n_k$. The natural CV analogue of one-bit teleportation leads to fault-tolerant implementations of gates in $\mathcal{D}^n_k$:
$$\begin{quantikz}
    \lstick{\ket{x=0}}&\gate{F}&\gate{D}&\ctrl{1}&\gate{DT_{-s}D^*}&D\ket{\psi} \\
    \lstick{\ket{\psi}}&\gate{F^2}&&\targ{}&\meterD{x\vphantom{0}}\wire[u][1]["x=s"]{c}
\end{quantikz}$$
where $F=R(\frac{\pi}{2})=e^{\frac{i\pi}{4}(Q^2+P^2)}$ is the Fourier transform  and $CNOT=e^{iQ_1P_2}$. This circuit allows for the off-line preparation of the magic resource state $D|p=0\rangle=DF|x=0\ra$ because of two properties of $D$: it is in the Clifford hierarchy, allowing it to be commuted backward through $T_{-s}$ leaving behind a fault-tolerantly-implementable gate in $\mathcal{C}_{k-1}^n$; and it is diagonal, so $D\otimes I$ commutes with $CNOT$.

For $n$-qudit systems with $d$ a prime power, a complete characterization of the diagonal elements of the Clifford hierarchy was obtained in \cite{CGK}. More specifically, when $n=1$ and $d$ is a prime, they are of the form
$$U=\sum_{j\in\Z_d}e^{2\pi i\sum_m p_m(j)/p^m}|j\ra\la j|,$$
where $p_m$ is a polynomial over $\Z_{d^m}$ and the level of the hierarchy depends on the largest value $m_*$ of $m$ in the above sum and the degree of $p_{m_*}$. 

We now establish an analogous characterization in the continuous variable setting, showing that all diagonal gates are exponentials of polynomials in the $n$-position operators $Q_1,...,Q_n$. Our argument involves new elements not present in the discrete setting of \cite{CGK}. 

\begin{theorem}\label{t:diag} For $n,k\in\N$, 
$$\mc{D}^n_k=\{e^{i p_k(Q_1,Q_2,...,Q_n)}\mid p_k \ \textnormal{multivariate polynomial of degree} \ \leq k\}.$$
\end{theorem}

The outline of our proof is to first show that any $U=M_u\in\mc{D}^n_k$ with $u\in L^\infty(\R^n)$ equals $M_v$ for a \textit{continuous} function $v:\R^n\to\T$. This is achieved by applying a recent automatic continuity result \cite[Theorem 1.6]{MS} to a measurable cocycle determined by the original function $u$. We can then apply some basic results in topology and distribution theory to show that $v=e^{i\theta(\cdot)}$ for a $C^1$-function $\theta:\R^n\to\R$, which is necessarily a polynomial of the correct degree by the induction hypothesis. We isolate the first part of the proof as a separate lemma, wherein $\lambda_a$ represents left translation by $a\in\R^n$.

\begin{lemma}\label{l:cts} Let $n\in\N$, and let $u\in L^\infty(\R^n)$ be a unitary such that for every $a\in\R^n$, there is a continuous function $g_a:\R^n\to\T$ such that $\lambda_a(u)\overline{u}=g_a$ a.e. Then there is a continuous $v:\R^n\to\T$ with $u=v$ a.e.
\end{lemma}

\begin{proof} First, $u\in L^\infty(\R^n)$ being unitary means there exists a (Lebesgue) measurable function (also denoted) $u$ in its equivalence class satisfying $|u(x)|=1$ for a.e. $x\in\R^n$. Letting $S$ denote the null set $\{x\in\R^n\mid |u(x)|\neq 1\}$, define $v:\R^n\to\T$ to be 1 on $S$ and equal to $u$ elsewhere. Then $v=u$ a.e., so is measurable (as Lebesgue measurable sets are complete). 

It follows by hypothesis that $g_a(x)=v(x-a)\overline{v(x)}$ for every $a\in\R^n$ and a.e. $x\in\R^n$. Put $g:\R^n\times\R^n\ni (a,x)\to g_a(x)\in\T$. Then for all $a,b\in\R^n$ and a.e. $x\in\R^n$,
\begin{align*} g(a+b,x)&=v(x-a-b)\overline{v(x)}\\
&=v(x-a-b)\overline{v(x-b)}v(x-b)\overline{v(x)}\\
&=g(a,x-b)g(b,x).
\end{align*}
Since the left and right hand sides are both continuous in $x$, (for fixed $a,b$), it follows that $g(a+b,x)=g(a,x-b)g(b,x)$ for all $a,b,x\in\R^n$. That is, $g$ is a $\T$-valued cocycle for the left translation action of $\R^n$ on itself, which is continuous in the second variable. Moreover, for every $x\in\R^n$, the function $a\mapsto g(a,x)$ coincides a.e. with the (Lebesgue) measurable function $a\mapsto v(x-a)\overline{v(x)}$ and is thus measurable. By \cite[Theorem 1.6]{MS}, $g:\R^n\times\R^n\to\T$ is (jointly) continuous. The equality
$$|v(x-a)-v(x)|=|(v(x-a)-v(x))\overline{v(x)}|=|g(a,x)-g(0,x)|, \ \ \ \forall \ a,x\in\R^n$$
then implies the continuity of $v$. 
\end{proof}

\begin{proof}[Proof of Theorem \ref{t:diag}] Fix $n\in\N$. Proposition \ref{p:inclusion} shows one containment, so it suffices to show that
$$\mc{D}^n_k\subseteq\{e^{i p_k(Q_1,Q_2,...,Q_n)}\mid p_k \ \textnormal{multivariate polynomial of degree} \ \leq k\}, \ \ \ \forall \ k\in\N,$$
which we prove by induction on $k$. For $k=1$, any $U\in\mc{D}^n_1$ is of the form $U=\pi(a,b,t)=e^{i t}e^{i ab}M_bT_a$ for some $a,b\in\R^n$ and $t\in\R$. But then $T_a=e^{-i t}e^{-i ab}M_{-b}U\in M(L^\infty(\R^n))$,
which forces $a=0$ as any diagonal translation operator is a scalar multiple of the identity. Then $U=e^{i t}M_b=e^{i(t1+bQ)}$, which is of the claimed form.

Assume the claim is true for $k-1$, and let $U\in\mc{D}^n_k$. Then for every $a,b\in\R^n$, $U_{a,b}:=U\pi(a,b)U^*\in\mc{C}^n_{k-1}$. But, as $U\in M(L^\infty(\R^n))$,
$$U_{a,b}=e^{i ab}M_bUT_aU^*=\pi(a,b)T_a^*UT_aU^*.$$
Now, $U=M_u$ for a unitary $u\in L^\infty(\R^n)$, and a standard calculation shows that $T_a^*UT_a=M_{\lambda_au}$. Since $\mc{C}^n_1\mc{C}^n_{k-1}\subseteq\mc{C}^n_{k-1}$, it follows that $M_{\lambda_a(u)\overline{u}}=\pi(a,b)^*U_{a,b}\in\mc{D}^n_{k-1}$. By the inductive hypothesis, for each $a\in\R^n$, $M_{\lambda_a(u)\overline{u}}=M_{g_a}$, where $g_a(x)=e^{i p_a(x)}$ for some polynomial $p_a$ of degree $\leq k-1$. Thus, $\lambda_a(u)\overline{u}=g_a$ a.e. Since $g_a:\R^n\to\T$ is continuous for every $a\in\R^n$, Lemma \ref{l:cts} produces a continuous $v:\R^n\to\T$ for which $u=v$ a.e. In particular, $U=M_v$, and $\lambda_a(v)\overline{v}=g_a$ for all $a\in\R^n$.

Next, as $e^{i(\cdot)}:\R\to\T$ is a covering map, $v:\R^n\to\T$ is continuous, $\T$ is path connected and $\R^n$ is simply connected, the standard lifting criterion implies the existence of a continuous $\theta:\R^n\to\R$ such that $v(x)=e^{i \theta(x)}$, $x\in\R^n$ (see, e.g., \cite[Corollary 6.4]{GH}). Then for every $a,x\in\R^n$, $\theta(x-a)-\theta(x)-p_a(x)=2\pi n_a(x)$ for some integer $n_a(x)\in\Z$. By continuity of the left hand side, $n_a(x)\equiv n_a$, so redefining $p_a(x)$ as $p_a(x)+2\pi n_a$, we may assume that for each $a\in\R^n$ $\lambda_a(\theta)-\theta=p_a$ is a polynomial of degree $\leq k-1$.

Now, as $\theta$ is continuous, it defines a distribution $T_\theta\in C_c^\infty(\R^n)'$ via integration:
$$\la T_\theta,f\ra=\int_{\R^n}\theta(x)f(x) \ dx, \ \ \ f\in C^\infty_c(\R^n).$$ 
For each $f\in C^\infty_c(\R^n)$ and $j=1,...,n$, it follows (by the mean value theorem and uniform continuity of $\partial_jf$) that $\partial_jf=\lim_{h\to0^+}h^{-1}(\lambda_{-he_j}f-f)$ uniformly. The same is true for $\partial_\alpha\partial_jf$, $\alpha\in\N_0^n$. Thus, the partial distributional derivative $\partial_jT_\theta$ satisfies
\begin{align*}\la\partial_jT_\theta,f\ra&=-\la T_\theta,\partial_jf\ra\\
&=-\lim_{h\to 0^+}\frac{1}{h}\int_{\R^n}\theta(x)(f(x+he_j)-f(x))\ dx\\
&=-\lim_{h\to0^+}\frac{1}{h}\int_{\R^n}(\theta(x-he_j)-\theta(x))f(x) \ dx\\
&=-\lim_{h\to0^+}\frac{1}{h}\int_{\R^n}p_{he_j}(x)f(x) \ dx\\
&=-\lim_{h\to0^+}\sum_{|\alpha|\leq k-1}\frac{1}{h}\int_{\R^n}c_{\alpha}(he_j)x^\alpha f(x) \ dx\\
&=-\lim_{h\to0^+}\bigg\la \sum_{|\alpha|\leq k-1}\frac{c_{\alpha}(he_j)}{h}T_{q_\alpha},f\bigg\ra,
\end{align*}
where $q_\alpha(x)=x^\alpha$. Since $f\in C^\infty_c(\R^n)$ was arbitrary, $\partial_jT_\theta$ belongs to the closure of the finite-dimensional subspace $\mathrm{span}\{T_{q_\alpha}\mid |\alpha|\leq k-1\}$ of the Hausdorff topological vector space $(C_c^\infty(\R^n)',\sigma(C_c^\infty(\R^n)',C_c^\infty(\R^n)))$. Thus $\partial_fT_\theta\in \mathrm{span}\{q_\alpha\mid |\alpha|\leq k-1\}$ as the latter space is closed \cite[Corollary 9.1.2]{Treves}. Since $\theta$ is continuous and $\partial_jT_\theta=\sum_{|\alpha|\leq k-1} c_{\alpha,j}q_\alpha$ coincides with a continuous function in the sense of distributions, \cite[Theorem 3.1.7]{Horm} implies that $\partial_j\theta(x)=\sum_{|\alpha|\leq k-1} c_{\alpha,j}q_\alpha(x)$, $x\in\R^n$, so that $\theta\in C^1(\R^n)$. The fundamental theorem of calculus applied to the line segment $\gamma:[0,1]\ni t\mapsto tx\in\R^n$ joining the origin to $x\in\R^n$ shows that 
$$\theta(x)=\theta(0)+\int_{0}^1\sum_{j=1}^n\sum_{|\alpha|\leq k-1}c_{\alpha,j}(tx)^\alpha x_j$$
is a polynomial of degree $\leq k$, which concludes the proof.
\end{proof}

\section{Semi-Clifford Gates}\label{s:semi-Clifford}

Qubit (and qudit) gates in 
$\mc{C}^n_2\mc{D}^n_k\mc{C}^n_2$ admit a reduction in gate teleportation resources based on the one-bit teleportation scheme for diagonal gates \cite{dS,ZCC}, validating their study for fault-tolerant quantum computing. The resulting subclass, known as \textit{semi-Clifford gates}, are equivalently defined as unitaries which send one maximal abelian subgroup of the Pauli group to another via conjugation. For $n=1,2$, all qubit and qudit ($d$ odd prime) hierarchy gates are semi-Clifford \cite{Aftab,CdS,dS,dSL,ZCC}, but when $k\ge4,n\ge3$, there exist gates that are not semi-Clifford \cite{ZCC}. It was conjectured that all third-level gates in any number of qubits are semi-Clifford \cite[Conjecture 1]{ZCC}, which was subsequently shown false by a counter-example of Gottesman-Mochon when $k=3$, $n=7$ (see \cite[\S1.1]{BS}).

If a CV gate is of the form $UDV\in\mc{C}^n_2\mc{D}^n_k\mc{C}^n_2$ we have an analogous one-bit teleportation scheme:

$$\begin{quantikz}
    \lstick{\ket{x=0}}&\gate{F}&\gate{D}&\ctrl{1}&\gate{DT_{-s}D^*}&\gate{U}&UDV\ket{\psi} \\
    \lstick{\ket{\psi}}&\gate{V}&\gate{F^2}&\targ{}&\meterD{x\vphantom{0}}\wire[u][1]["x=s"]{c}
\end{quantikz}$$

Since $U\in\mc{C}^n_2$, one can also commute it backward ahead of the measurement at the expense of a (potentially) different conditional $\mc{C}^{n}_{k-1}$-gate:

$$\begin{quantikz}
    \lstick{\ket{x=0}}&\gate{F}&\gate{D}&\ctrl{1}&\gate{U}&\gate{UDT_{-s}D^*U^*}&UDV\ket{\psi} \\
    \lstick{\ket{\psi}}&\gate{V}&\gate{F^2}&\targ{}&&\meterD{x\vphantom{0}}\wire[u][1]["x=s"]{c}
\end{quantikz}$$

Hence the analogous CV subhierarchy of semi-Clifford gates is also of interest for fault-tolerant schemes. In this section, we introduce this subhierarchy and show that, contrary to the qubit setting, the (analogue of the) Zeng-Chen-Chuang conjecture is true: for any $n$-mode system, all third-level gates $\mc{C}^n_3$ are semi-Clifford (Theorem \ref{t:SC}).

Recall that a subspace $S\leq\R^{2n}$ is \textit{isotropic} if $\Om(s_1,s_2)=0$ for all $s_1,s_2\in S$, and \textit{Lagrangian} if it is maximally isotropic. 
A unitary $U\in\mc{U}(L^2(\R^n))$ is \textit{semi-Clifford} if there exist Lagrangian subspaces $L_1,L_2\leq\R^{2n}$ for which $U\pi(L_1\times\R)U^*=\pi(L_2\times\R)$. The following Proposition justifies our definition, and is a continuous variable analogue of \cite[Proposition II.1]{ZCC}. 

\begin{proposition}\label{p:SC}
If $U\in\mc{U}(L^2(\R^n))$ is semi-Clifford there exist $C_1,C_2\in\mc{C}^n_2$ such that $C_1UC_2\in M(L^\infty(\mathbb{R}^n))$.
\end{proposition}

\begin{proof}
Let $L_1$ and $L_2$ be Lagrangian subspaces for which $U\pi(L_1\times\R)U^*=\pi(L_2\times\R)$. By \cite[Lemma 2.17]{Folland}, there exists $S_2,S_3\in Sp(2n,\R)$ such that $\pi(L_1\times\R)=\mu(S_2)\pi(\{0\}\times\mathbb{R}^n\times \R)\mu(S_2)^*$ and $\pi(L_2\times\R)=\mu(S_3)^*\pi(\{0\}\times\mathbb{R}^n\times\R)\mu(S_3)$. Then $$U\mu(S_2)\pi(\{0\}\times\mathbb{R}^n\times\R)\mu(S_2)^*U^*=\mu(S_3)^*\pi(\{0\}\times\mathbb{R}^n\times\R)\mu(S_3),$$
so $\pi(\{0\}\times\mathbb{R}^n\times\R)=\mu(S_3)U\mu(S_2)\pi(\{0\}\times\mathbb{R}^n)\mu(S_2)^*U^*\mu(S_3)^*$.

Let $V=\mu(S_3)U\mu(S_2)$. Then for any $b\in\mathbb{R}^n$, \begin{equation}\label{e:V}VM_bV^*=V\pi(0,b,0)V^*=\pi(0,b',t_b)\end{equation}
for some $b'\in\mathbb{R}^n$, $t_b\in\R$. This defines a homomorphism $\R^n\ni b\mapsto (b',e^{it_b})\in\R^n\times\T$, which is also continuous: if $b_j\to b$ in $\R^n$, then $$\pi(0,b_j',t_{b_j})=VM_{b_j}V^*\to VM_bV^*=\pi(0,b',t_b)$$
in the strong operator topology. Since $\pi(H_n)\cong\R^{n}\times\R^{n}\times\T$ is a topological isomorphism, we have $(b'_j,e^{it_{b_j}})\to(b',t_b)$ in $\R^n\times\T$.
Thus, $(b',e^{it_b})=(Ab,e^{iab})$ for some $A\in M_n(\R)$ and $a\in\R^n$. By the unitary implementation (\ref{e:V}), if $Ab=0$  then
$$M_b=V^*\pi(0,Ab,ab)V=e^{iab}1,$$
forcing $b=0$. Thus, $A$ is invertible. Let $$D_A=\left[\begin{array}{cc}
    (A^t)^{-1} & 0 \\
    0 & A
\end{array}\right]\in Sp(2n,\mathbb{R})$$
and $C_1=T_a^*\mu(D_A)^*\mu(S_3)\in\mc{C}^n_2$. Then $$VM_bV^*=e^{iab}\pi(D_A(0,b))=\mu(D_A)T_aM_bT_a^*\mu(D_A)^*$$
so that  $$(C_1U\mu(S_2))M_b(\mu(S_2)^*UC_1^*)=T_a^*\mu(D_A)^*VM_bV^*\mu(D_A)T_a=M_b, \ \ \  b\in\mathbb{R}^n.$$
Since the $\mathrm{span}\{M_b\mid b\in\R^n\}$ is weak*-dense in $L^\infty(\R^n)$ (combine \cite[Theorem 3.27]{Folland2} with \cite[Proposition 3.33]{Folland2}), it follows that $C_1U\mu(S_2)M_f=M_fC_1U\mu(S_2)$ $\forall f\in L^\infty(\mathbb{R}^n)$. Thus $C_1U\mu(S_2)\in M(L^\infty(\mathbb{R}^n))'=M(L^\infty(\R^n))$ \cite{KR}.
\end{proof}

\begin{remark} The maximal abelian subgroups of $\mc{C}^n_1=\pi(H_n)$ are of the form $\pi(S\times\R)$, where $S$ is maximally integral-isotropic in the sense that it is a maximal subgroup satisfying $\Omega(s_1,s_2)\in 2\pi\Z$ for all $s_1,s_2\in S$ (e.g., $S=S_k\times 2\pi\Z^{2(n-k)}$, with $0\leq k\leq n$ and $S_k\leq\R^{2k}$ isotropic). It would be natural to study the corresponding set of semi-Clifford gates. We have restricted to a ``purely continuous'' notion in this first paper.
\end{remark}

Given $n,k\in\N$, we let $\mc{SC}_k^n$ denote the set of semi-Clifford gates in $\mc{C}_k^n$. 

\begin{proposition}\label{p:decomp} For every $n,k\in\N$, $\mc{SC}_k^n=\mc{C}^n_2\mc{D}^n_k\mc{C}^n_2$.
\end{proposition}

\begin{proof} Combining Theorem \ref{t:diag}, Proposition \ref{p:SC} and the Clifford group invariance $\mc{C}^n_2\mc{C}^n_k\mc{C}^n_2\subseteq\mc{C}^n_k$ shows that $\mc{SC}_k^n\subseteq\mc{C}^n_2\mc{D}^n_k\mc{C}^n_2$.

Conversely, if $U=C_1DC_2$ for some $C_1,C_2\in\mc{C}^n_2$ and $D\in\mc{D}^n_k$. Then $U\in\mc{C}^n_k$ and $C_j=\mu(S_j)\pi(a_j,b_j,t_j)$, $j=1,2$. The subspaces $L_1=S_1(\{0\}\times\R^n)$ and $L_2=S_2^{-1}(\{0\}\times\R^n)$ are Lagrangian and $U\pi(L_2\times\R)U^*=\pi(L_1\times\R)$.
\end{proof}

We now show that all third level gates are semi-Clifford. The proof naturally reduces to the structure of Lie algebra homomorphisms $\mathfrak{h}_n\to sp(2n,\R)\ltimes\mathfrak{h}_n$ which fix the center of $\mathfrak{h}_n$. The $sp(2n,\R)$-component of any such homomorphism is necessarily 2-nilpotent, and therefore strictly upper triangular up to conjugation. This triangularity gives rise to a Lagrangian subspace leading to the semi-Clifford property.

\begin{theorem}\label{t:SC} For every $n\in\N$, 
$\mathcal{C}_3^n=\mc{SC}_3^n$.
\end{theorem}

\begin{proof}
Given $U\in \mathcal{C}_3^n$, the composition
$$\pi_U:=\mathrm{Ad}(U)\circ\pi:H_n\to\mathcal{C}^n_2\cong  \widetilde{HSp(2n,\R)}/\mathrm{Ker}(\tilde{\om})$$
is a smooth homomorphism. Since $\tilde{\om}:\widetilde{HSp(2n,\R)}\to\mc{C}_2^n$ is a smooth homomorphism with discrete kernel (combine \cite[pg 196]{Folland} with the universal covering property), it is a covering map and the Lie algebra of 
$$\widetilde{HSp(2n,\R)}/\mathrm{Ker}(\tilde{\om})=\mathrm{Lie}(HSp(2n,\R))= sp(2n,\R)\ltimes \mathfrak{h}_n.$$
Simple (and local path) connectedness of $H_n$ yields a unique smooth lift $\vphi:H_n\to \widetilde{HSp(2n,\R)}$ satisfying $\pi_U=\tilde{\om}\circ \vphi$ and $\vphi(e)=e$ \cite[Corollary 6.4]{GH}. By uniqueness, $\vphi$ is automatically a homomorphism (as $g\mapsto \vphi(gg_0)\vphi(g_0)^{-1}$ is another such lift). Moreover, the subspace of smooth vectors for $\pi_U$, $\mathcal{H}^\infty(\pi_U)=U\mc{S}(\R^n)\supseteq\mc{H}^\infty(\tilde{\om})\supseteq\mc{S}(\R^n)$.

On $\mc{S}(\R^n)$, the differential $d\pi_U$ satisfies 
$$d\pi_U(v)=Ud\pi(v)U^*=d\tilde{\om}(d\vphi(v)), \ \ \ v\in\R^{2n},$$
where $d\vphi:\mathfrak{h}_n\to sp(2n,\R)\ltimes \mathfrak{h}_n$ is a Lie algebra homomorphism.  Viewing $sp(2n,\R)\ltimes \mathfrak{h}_n$ as a direct sum of its subspaces of polynomials of degree two, one, and zero in $Q$ and $P$, we can write $$d\vphi(v)=R_2(v)+R_1(v)+R_0(v),$$
where $R_2(v)\in sp(2n,\R)$, $R_1(v)\in\R^{2n}$, and $R_0(v)\in\R1$. Note that each $R_k(v)$ is linear in $v\in\R^{2n}$ (composition of $d\vphi$ with a linear projection). For arbitrary $v,w\in\mathbb{R}^{2n}$ we compute the commutator
\begin{align*}
-i\Omega(v,w)1&=[U(v(P,Q))U^*,U(w(P,Q))U^*]\\
&=[d\pi_U(v),d\pi_U(w)]\\
&=d\tilde{\om}[R_1(v)+R_2(v),R_1(w)+R_2(w)]\\
&=d\tilde{\om}([R_2(v),R_2(w)]+[R_1(v),R_2(w)]\\
&+[R_2(v),R_1(w)]+[R_1(v),R_1(w)]).
\end{align*}
Comparing the degrees (as polynomials in $P$ and $Q$) of both sides and appealing to injectivity of $d\tilde{\om}$, it follows that
\begin{align*}
    &(a) \ \ [R_2(v),R_2(w)]=0,\\
    &(b) \ \ [R_1(v),R_2(w)]+[R_2(v),R_1(w)]=0,\\
    &(c) \ \ [R_1(v),R_1(w)]=-\Omega(v,w).
\end{align*}
Relation $(c)$ implies that $R_1(\cdot):\R^{2n}\to\R^{2n}$ is symplectic, meaning $R_1(v)=S_1v$, $v\in\R^{2n}$, for some $S_1\in Sp(2n,\mathbb{R})$. 
By nondegeneracy of $\Omega$, we can write $R_0(v)=\Omega(-r_0,v)$ for some $r_0\in\mathbb{R}^{2n}$. Then $\pi(r_0)d\pi(v)\pi(r_0)^*=d\pi(v)+\Omega(r_0,v)1$, so that
\begin{equation}\label{e:U}U\pi(r_0)d\pi(v)\pi(r_0)^*U^*=Ud\pi(v)U^*+\Omega(r_0,v)1=d\tilde{\om}(R_2(v)+S_1v)\end{equation}
for all $v\in\mathbb{R}^{2n}$.
By the Lie bracket structure on the semi-direct product $sp(2n,\R)\ltimes \mathfrak{h}_n$, we have $$[S_1v,R_2(w)]=(0,-R_2(w)S_1v), \ \ [R_2(v),S_1w]=(0,R_2(v)S_1w), \ \ \ v,w\in\R^{2n}.$$ 
Thus, by $(b)$, we have $R_2(v)S_1w=R_2(w)S_1v$. Letting $N(v):=S_1^{-1}R_2(v)S_1$, the family $N(\R^{2n})$ satisfies the symmetry condition $N(v)w=N(w)v$, and by $(a)$, $[N(v),N(w)]=0$ for all $v,w\in\R^{2n}$. 

We now show that $N(v)N(w)=0$ for all $v,w\in\R^{2n}$, which will allow us to give a simultaneous strictly upper triangular representation of each $N(v)$. Consider the multi-linear form
$$p(v,w,x,y)=\Omega(N(v)N(w)x,y), \ \ \ \ v,w,x,y\in\R^{2n}.$$
By the symmetry condition and commutativity, $p$ is $S_3$-symmetric in its first 3 arguments. Moreover, by symplecticity,

$$p(v,w,x,y)=-\Omega(N(w)x,N(v)y)=-\Omega(N(w)x,N(y)v)=p(y,w,x,v),$$
so that $p$ is invariant under the permutation group $S_3$ and the transposition $(14)$. As $S_3$ and $(14)$ generate all of $S_4$, $p$ is totally symmetric. However, by anti-symmetry of $\Omega$, 
$$p(v,w,x,y)=-\Omega(y,N(v)N(w)x)=-\Omega(N(w)N(v)y,x)=-p(v,w,y,x)=-p(v,w,x,y),$$
so we must have $p(v,w,x,y)=0$ for all $v,w,x,y\in\R^{2n}$. Equivalently, $N(v)N(w)=0$ for all $v,w\in\R^{2n}$.

The subspace $L_0=\mathrm{span}\{\mathrm{Im}(N(v))\mid v\in\R^{2n}\}$ is then isotropic as
$$\Omega(N(v)x,N(w)y)=-p(w,v,z,y)=0, \ \ \ x,y,v,w\in\R^{2n}.$$
Let $L$ be a Lagrangian subspace for which $L_0\subseteq L$, and let $S_2\in Sp(2n,\R)$ be such that $L=S_2^{-1}(\R^n\times\{0\})$. As $N(v)(\R^{2n})\subseteq L_0\subseteq S_2^{-1}(\R^n\times\{0\})$, it follows that
$S_2N(v)S_2^{-1}(\R^{2n})\subseteq\R^n\times\{0\}$, so we must have the following block structure
$$S_2N(v)S_2^{-1}=\begin{bmatrix}A(v) & M(v)\\0 & 0\end{bmatrix}.$$
However, $S_2N(v)S_2^{-1}\in sp(2n,\R)$, which forces $A(v)=0$ and $M(v)^t=M(v)$. Then, if $u\in L$, $S_2u=(a,0)$ for some $a\in\R^n$, in which case
\begin{align*}\bigg\la\begin{bmatrix}0&0 \\ 0&-M(u) \end{bmatrix}S_2v,S_2w\bigg\ra&=\Omega(S_2N(u)S_2^{-1}(S_2v),S_2w)\\
&=\Omega(N(u)v,w)\\
&=\Omega(N(v)u,w)\\
&=\Omega(S_2N(v)S_2^{-1}(S_2u),S_2w)\\
&=\bigg\la\begin{bmatrix}0&0 \\ 0&-M(v) \end{bmatrix}\begin{bmatrix}a\\0\end{bmatrix},Tw\bigg\ra\\
&=0.
\end{align*}
Since the above equalities hold for all $v,w\in\R^{2n}$ and $S_2$ is bijective, we must have $M(u)=0$. Thus, $L\subseteq\mathrm{Ker}(M)=\mathrm{Ker}(N)=\mathrm{Ker}(R_2)$, so for $v\in L$, equation (\ref{e:U}) shows that 
$$U\pi(r_0)d\pi(v)\pi(r_0)^*U^*=d\tilde{\om}(R_2(v)+S_1v)=d\pi(S_1v),$$
as $\tilde{\om}|_{H_n}=\pi$. Exponentiation shows that $U\pi(r_0)$ conjugates $\pi(L\times\R)$ onto $\pi(S_1L\times\R)$.
\end{proof}

Inspection of the previous argument leads to the following classification of $\mc{C}_3^n$ gates up to phase.

\begin{theorem}\label{t:classify} Let $n\in\N$. There is a one-to-one correspondence between $\mc{C}^n_3/\T$ and triples $(T,S,r)$ where 
\begin{itemize}
\item $T:\R^{2n}\times\R^{2n}\times\R^{2n}\to\R$ is a symmetric tri-linear form for which there is an isotropic subspace $L_0\leq\R^{2n}$ such that $T(u,v,w)=0$ whenever one of its arguments belongs to $L_0^\Omega$ (in short, $T$ has isotropic support). 
\item $S\in Sp(2n,\R)$;
\item and $r\in\R^{2n}$.
\end{itemize}
\end{theorem}

\begin{proof} Let $U\in\mc{C}_3^n$. Then, using notation from the proof of Theorem \ref{t:SC}, the map
$$T(u,v,w)=\Omega(N(u)v,w), \ \ \ u,v,w\in\R^{2n},$$
is a symmetric tri-linear form such that $T(u,v,w)=0$ whenever $w\in L_0^{\Omega}$, where $L_0:=\mathrm{span}\{\mathrm{Im}(N(v))\mid v\in\R^{2n}\}$. Taking $S$ to be the symplectic matrix satisfying $R_2(v)=SN(v)S^{-1}$, and $R_1(v)=Sv$, and letting $r\in\R^{2n}$ satisfy $R_0(v)=rv$, it follows that the assignment $\mc{C}_3^n/\T\ni \mathrm{Ad}(U)\mapsto (T,S,r)$ is injective. 

Next, let $(T,S,r)$ be a triple with the stated properties. Out of this data, we first build a Lie algebra homomorphism $\vphi:\mathfrak{h}_n\to sp(2n,\R)\ltimes \mathfrak{h}_n$. 

Since $T$ is a symmetric tri-linear form, for fixed $u\in\R^{2n}$, the symmetric bi-linear form $T(u,\cdot,\cdot)$ can be represented as
$$T(u,v,w)=\Omega(N(u)v,w), \ \ \ v,w\in\R^{2n},$$
for some $N(u)\in sp(2n,\R)$. The resulting map $N:\R^{2n}\to sp(2n,\R)$ is linear. Let $L_0$ be the isotropic support of $T$. Then, for any $w\in L_0^{\Omega}$, 
$$0=T(u,v,w)=\Omega(N(u)v,w), \ \ \ u,v\in\R^{2n}.$$
Thus, $N(u)v\in L_0$ for all $u,v\in\R^{2n}$, implying that
$$\Omega(N(u_1)N(u_2)v,w)=-\Omega(N(u_2)v,N(u_1)w)=0, \ \ \ u_1,u_2,v,w\in\R^{2n}.$$
Hence, $N(u)N(v)=0$ for all $u,v\in\R^{2n}$. Moreover, symmetry of $T$ entails $N(u)v=N(v)u$, $u,v\in\R^{2n}$.

Define $\vphi:\mathfrak{h}_n\to sp(2n,\R)\ltimes \mathfrak{h}_n$ by
$$\vphi(v,t)=SN(v)S^{-1} +Sv +rv+t.$$
Then $\vphi$ is an injective linear map which fixes the center of $\mathfrak{h}_n$. By the commutation relations in the semi-direct product Lie algebra $sp(2n,\R)\ltimes \mathfrak{h}_n$, together with the 2-nilpotency ($N(u)N(v)=0$) and symmetry ($N(u)v=N(v)u$) conditions, we have
$$[\vphi(u,s),\vphi(v,t)]=SN(u)v-SN(v)u+\Omega(Su,Sv)=\Omega(u,v)=[(u,s),(v,t)]$$
for all $(u,s),(v,t)\in\mathfrak{h}_n$. Thus, $\vphi$ is a Lie algebra homomorphism. 

Similar to the proof of the previous theorem, there is $S_2\in Sp(2n,\R)$ for which
$$S_2N(v)S_2^{-1}=\begin{bmatrix}0 & 0\\M(v) & 0\end{bmatrix}$$
for a symmetric $M(v)$ such that $S_2^{-1}(\{0\}\times\R^n)\subseteq\mathrm{Ker}(M)$. Then $M(v)=M(S_2^{-1}((S_2v)_P,0))$ and the differentiated metaplectic representation (\ref{e:quadratic}) then satisfies
$$d\mu(S_2N(v)S_2^{-1})\psi(x)=-\frac{i}{2}xM(v)x\psi(x), \ \ \ \psi\in\mc{S}(\R^n).$$
Conjugating $\vphi$ with $S_2S^{-1}$ yields an equivalent Lie algebra homomorphism $$\vphi_2(v,t)=S_2N(v)S_2^{-1}+S_2v+rv+t.$$
Composing with the differentiated extended metaplectic representation yields the Lie algebra representation
$$(v,t)\mapsto i(-\frac{1}{2}QM(v)Q+S_2v(P,Q)+(rv+t)1).$$
By simple connectedness of $H_n$, this integrates to a smooth Lie group representation $\pi_2:H_n\to\mc{U}(L^2(\R^n))$ satisfying 
$$\pi_2(v,t)=e^{i(rv+t)}e^{i(-\frac{1}{2}QM(v)Q+S_2v(P,Q))}.$$
Note that the directional derivative of $T(u,u,u)$ in the direction $v$ is 
$$v\nabla T(u,u,u)=3\Omega(N(v)u,u).$$
It follows that for $x\in\R^n$, 
$$xM(v)x=\frac{1}{3}v\nabla T(S_2^{-1}(x,0),S_2^{-1}(x,0),S_2^{-1}(x,0)).$$
Hence, there is a degree-$3$ polynomial $p_3:\R^n\to\R$ for which $$q\nabla p_3(x)=-\frac{1}{2}xM(S_2^{-1}(q,0))x, \ \ \ q,x\in\R^n.$$
Then, using the commutation relations (\ref{e:dirder}) from the proof of Proposition \ref{p:inclusion}, it follows that
$$
e^{ip_3(Q)}(S_2v(P,Q))e^{-ip_3(Q)}=S_2v(P,Q)+(S_2v)_P\nabla p_3(Q)=S_2v(P,Q)-\frac{1}{2}QM(v)Q.$$
Thus,
$$\pi_2(v,t)=e^{irv}e^{ip_3(Q)}\pi(S_2v(P,Q),t)e^{-ip_3(Q)}=e^{irv}e^{ip_3(Q)}\mu(S_2)\pi(v,t)\mu(S_2)^*e^{-ip_3(Q)}.$$
It follows that $\pi_2$ and hence the integrated representation of $\vphi$ is equivalent via a unitary in $\mc{C}^n_3$ to the Schr\"{o}dinger representation, which completes the proof. 
\end{proof}

One can produce examples of non-semi-Clifford gates on 3 modes by the continuous variable analogue of an example of Zeng-Chen-Chuang \cite{ZCC}.

\begin{example} Let $U\in\mc{U}(L^2(\R^3))$ be the Toffoli gate
$$U\psi(x_1,x_2,x_3)=\psi(x_1,x_2,x_3-x_1x_2), \ \ \ x_1,x_2,x_3\in\R, \ \psi\in L^2(\R^{3}),$$
and for $k\geq1$, let $V_k=e^{iQ_3^k}$. If $W_k=V_kU$, then standard (albeit lengthy) calculations show that $W_k\in\mc{C}^3_{k+1}$.

We now show $W_3\notin\mc{SC}^3_4$. To that end, first observe that $U=Fe^{iQ_1Q_2Q_3}F^*=e^{-iQ_1Q_2P_3}$, where $F$ is the Fourier transform on the third mode ($FQ_3F^*=-P_3$ in our convention). Then, for $a,b\in\R^3$, 
\begin{align*}W_3\pi(a,b)W_3^*&=V_3Fe^{iQ_1Q_2Q_3}F^*\pi(a,b)F e^{-iQ_1Q_2Q_3}F^*V_3^*\\
&=V_3Fe^{iQ_1Q_2Q_3}\pi(p,q)e^{-iQ_1Q_2Q_3}F^*V_3^* \ \ \ (\textnormal{$(p,q)=(a_1,a_2,b_3,b_1,b_2,-a_3)$})\\
&=V_3Fe^{iQ_1Q_2Q_3}e^{-i(Q_1+a_1)(Q_2+a_21)(Q_3+b_31)}\pi(p,q)F^*V_3^*\\
&=V_3e^{-iQ_1Q_2P_3}e^{i(Q_1+a_1)(Q_2+a_21)(P_3-b_31)}\pi(a,b)V_3^*,
\end{align*}

Supposing that $W_3$ is semi-Clifford, there exist Lagrangian subspaces $L_1,L_2\leq\R^6$ such that $W_3\pi(L_1\times\R)W_3^*=\pi(L_2\times\R)$. Let $(a,b)\in L_1$ be non-zero. Then there exist a (smooth) curve $\R\ni t\mapsto (r(t),s(t),z(t))\in L_2\times \R$ for which
$$\pi(r(t),s(t),z(t))=W_3\pi(ta,tb)W_3^*=V_3e^{-iQ_1Q_2P_3}e^{i(Q_1+ta_1)(Q_2+ta_21)(P_3-tb_31)}\pi(ta,tb)V_3^*, \ \ \ t\in\R.$$
Then
\begin{align*}e^{-iQ_3^3+i(Q_3+r_3(t)1)^3}\pi(r(t),s(t),z(t))&=V_3^*\pi(r(t),s(t),z(t))V_3\\
&=e^{-iQ_1Q_2P_3}e^{i(Q_1+ta_1)(Q_2+ta_21)(P_3-tb_31)}\pi(ta,tb),\end{align*}
so that
$$e^{i(Q_3^3-(Q_3+r_3(t)1)^3)}e^{-iQ_1Q_2P_3}e^{i(Q_1+ta_1)(Q_2+ta_21)(P_3-tb_31)}=e^{i(z(t)-\frac{t}{2}(r(t)b-s(t)a))}\pi(r(t)-ta,s(t)-tb).$$
Since the first 2 legs of the left hand side live in $M(L^\infty(\R))$ (i.e., are diagonal), we must have $r_1(t)=ta_1$ and $r_2(t)=ta_2$. Then
\begin{align*}&e^{-iQ_1Q_2P_3}e^{i(Q_1+ta_1)(Q_2+ta_21)(P_3-tb_31)}\\
&=e^{i\theta(t)}e^{i((Q_3+r_3(t)1)^3-Q_3^3)}e^{i((r_3(t)-ta_3)P_3+(s(t)-b)Q)},
\end{align*}
where $\theta(t)=(z(t)-\frac{t}{2}(r(t)b-s(t)a)) +\frac{1}{2}(r_3(t)-ta_3)(s_3(t)-tb_3)$. Since the operator exponents $i((Q_3+r_3(t)1)^3-Q_3^3)$ and $i((r_3(t)-ta_3)P_3+(s(t)-b)Q)$ belong to the nilpotent Lie algebra $\mathfrak{g}_3$ from the proof of Proposition \ref{p:inclusion}, we can apply the Baker-Campbell-Hausdorff formula (without convergence issues) to write
\begin{align*}&e^{i\theta(t)}e^{i((Q_3+r_3(t)1)^3-Q_3^3)}e^{i((r_3(t)-ta_3)P_3+(s(t)-b)Q)}\\
&=e^{i\theta(t)}e^{i(3r_3(t)Q_3^2+3ta_3r_3(t)Q_3+(r_3(t)-ta_3)P_3+(s(t)-tb)Q+\frac{t}{2}a_3r_3(t)(r_3(t)+ta_3)1)}.\end{align*}
On the other hand,
\begin{align*}&e^{-iQ_1Q_2P_3}e^{i(Q_1+ta_1)(Q_2+ta_21)(P_3-tb_31)}=e^{-i(Q_1Q_2P_3-(Q_1+ta_1)(Q_2+ta_21)(P_3-tb_31))}\\
&=e^{-i(tb_3Q_1Q_2+tb_2Q_1P_3+tb_1Q_2P_3-t^2b_2b_3Q_1-t^2b_1b_3Q_2-t^2b_1b_2P_3+t^3b_1b_2b_31)}.
\end{align*}
Equating the derivatives of the previous equations at $t=0$ yields
$$\theta'(0)1+3r_3'(0)Q_3^2+(r_3'(0)-a_3)P_3+(s'(0)-b)Q=-(b_3Q_1Q_2+b_2Q_1P_3+b_1Q_2P_3).$$
\end{example}
Linear independence implies that $r_3'(0)=a_3=0$ and $s'(0)=b=0$. Hence,
$$(a,b)=(a_1,a_2,0,0,0,0),$$
and since $(a,b)\in L_1$ was an arbitrary  $\mathrm{dim}(L_1)\leq 2$ -- contradiction. 

\section{The Smooth Hierarchy}\label{s:smooth}

Any gate $D\in\mc{D}_k^n$ in the diagonal hierarchy leaves the Schwartz space $\mc{S}(\R^n)$ invariant: by Theorem \ref{t:diag}, $D=e^{ i p_k(Q_1,...,Q_n)}$ for a degree-$k$ polynomial $p_k$, which, as a function of $q_1,...,q_n$, has polynomially bounded partial derivatives of all orders. Hence, $D\psi\in \mc{S}(\R^n)$ for all $\psi\in\mc{S}(\R^n)$. By \cite[Lemma 2.14]{Folland}, $\mc{C}^n_2\mc{S}(\R^n)\subseteq \mc{S}(\R^n)$, so it follows that any semi-Clifford gate $U\in\mc{SC}_k^n$ satisfies $U\mc{S}(\R^n)\subseteq\mc{S}(\R^n)$ and $U^*\mc{S}(\R^n)\subseteq\mc{S}(\R^n)$. Equivalently, $U\mc{S}(\R^n)=\mc{S}(\R^n)$. 

It is not clear to the authors whether the same property holds for any Clifford gate $U\in\mc{C}_k^n$ ($k\geq4$). As a natural and desirable property for a continuous variable gate to satisfy, we call the corresponding subset the \textit{smooth Clifford hierarchy}:
$$\mc{C}^{\infty,n}_k:=\{U\in\mc{C}_k^n\mid U\mc{S}(\R^n)=\mc{S}(\R^n)\}.$$
Clearly, $\mc{C}^{\infty,n}_k\subseteq \mc{C}^{\infty,n}_{k+1}$ for all $k,n\in\N$, and by Theorem \ref{t:SC}, $\mc{C}^{\infty,n}_k=\mc{C}^{n}_{k}$ for $k=1,2,3$ and $n\in\N$. By the closed graph theorem, any $U\in\mc{C}^{\infty,n}_k$ induces a continuous linear map on $\mc{S}(\R^n)$ with respect to its Fr\'{e}chet topology.

A simple, but useful reformulation which justifies our terminology is the following.

\begin{proposition}\label{p:smoothdef} For any $k,n\in\N$, $\mc{C}^{\infty,n}_k:=\{U\in\mc{C}_k^n\mid \mc{S}(\R^n)=\mc{H}^\infty(U\pi(\cdot)U^*)\}$, where $\mc{H}^\infty(U\pi(\cdot)U^*)$ denotes the space of smooth vectors for the representation $U\pi(\cdot)U^*$ of $H_n$.
\end{proposition}

\begin{proof} Simply combine the well-known fact that $\mc{H}^\infty(\pi)=\mc{S}(\R^n)$ (see, e.g., \cite[Theorem 2.2]{GT} and the references therein), together with the general relation  $\mc{H}^\infty(U\pi(\cdot)U^*)=U\mc{H}^\infty(\pi)$ for any unitary $U\in\mc{U}(L^2(\R^n))$.
\end{proof}

For $n,k\in\N$, let $\mc{A}^n_k$ denote the subspace of continuous linear operators on $\mc{S}(\R^n)$ given by $\C$-coefficient polynomials of degree $\leq k$ in $Q_1,...,Q_n,P_1,...,P_n$. Let $\mc{A}^n=\bigcup_{k\geq1}\mc{A}^n_k$ denote the complex Weyl algebra of all ($\C$-coefficient) polynomials in $Q_1,...,Q_n, P_1,...,P_n$. Gates in the smooth hierarchy behave as one might expect when acting infinitesimally on the canonical generators, as we now show.

\begin{proposition}\label{p:smooth} For every $n\in\N$, $k\geq 2$ and $U\in\mc{C}^{\infty,n}_k$, $U\mc{A}^n_1U^*\subseteq\mc{A}^n_{k-1}$.
\end{proposition}

\begin{proof} Fix $n\in\N$. We prove the claim by induction over $k\geq 2$. The base case ($k=2$) follows by definition of the Clifford group, as any $U\in\mc{C}^{\infty,n}_2=\mc{C}^n_2$ induces (by conjugation) an automorphism of $\mc{C}^n_1=\pi(H_n)$, and differentiation yields $$U\mc{A}_1^nU^*=U(d\pi(\mathfrak{h}_n)+id\pi(\mathfrak{h}_n))U^*\subseteq d\pi(\mathfrak{h}_n)+id\pi(\mathfrak{h}_n)=\mc{A}_1^n.$$

Suppose the claim holds for $k$, and let $U\in\mc{C}^{\infty,n}_{k+1}$. Then $U(g):=U\pi(g)U^*\in \mc{C}^{\infty,n}_{k}$ for all $g\in H_n$. Given  $Y,Z\in\mathfrak{h}_n$, $U(e^{tZ})d\pi(Y)U(e^{tZ})^*\in\mc{A}_{k-1}^n$ for all $t\in\R$ by assumption. As a finite-dimensional subspace of the locally convex Hausdorff space $\mc{L}(\mc{S}(\R^n))$ (equipped with the strong operator topology), $\mc{A}_{k-1}^n$ is closed \cite[Corollary 9.1.2]{Treves}. For $\psi\in\mc{S}(\R^n)$, $U^*\psi\in\mc{S}(\R^n)=\mc{H}^\infty(\pi)$, so by \cite[Proposition 10.1.11]{Schmudgen} the function $\R\ni t\mapsto \pi(e^{tZ})U^*\psi\in\mc{S}(\R^n)$ is smooth with respect to the Fr\'{e}chet topology. Since $U$ is continuous on $\mc{S}(\R^n)$, $\R\ni t\mapsto U(e^{tZ})\psi=U\pi(e^{tZ})U^*\psi\in\mc{S}(\R^n)$ is smooth and the derivative
$$\frac{d}{dt}U(e^{-tZ})\psi|_{t=0}=\lim_{t\to0}\frac{U(e^{-tZ})\psi-\psi}{t}=-Ud\pi(Z)U^*\psi$$
converges in $\mc{S}(\R^n)$. Since $d\pi(Y)$ is continuous on $\mc{S}(\R^n)$, it follows that
$$\frac{d}{dt}U(e^{tZ})d\pi(Y)U(e^{-tZ})\psi|_{t=0}=[Ud\pi(Z)U^*,d\pi(Y)]\psi.$$
Since the derivative converges in $\mc{S}(\R^n)$ and $\psi\in\mc{S}(\R^n)$ was arbitrary, it follows that $[Ud\pi(Z)U^*,d\pi(Y)]\in\mc{A}^n_{k-1}$ (as $\mc{A}^n_{k-1}$ is closed). The derivation $\ad_{Ud\pi(Z)U^*}=[Ud\pi(Z)U^*,\cdot]$ then satisfies $\ad_{Ud\pi(Z)U^*}(\mc{A}^n_1)\subseteq\mc{A}^n_{k-1}$. By repeated application of the Leibniz rule 
$$[Ud\pi(Z)U^*,AB]=[Ud\pi(Z)U^*,A]B+A[Ud\pi(Z)U^*,B],$$
(viewing $Ud\pi(Z)U^*\in\mc{L}(\mc{S}(\R^n))$ to make sense of the commutator), it follows that the derivation $\ad_{Ud\pi(Z)U^*}(\mc{A}^n)\subset\mc{A}^n$. Since derivations of the Weyl algebra $\mc{A}^n$ are inner \cite{Dix66}, there is some $H\in\mc{A}^n$ such that $Ud\pi(Z)U^*=H+\lambda1$, $\lambda\in\C$. 

If $H\in\C1$, there is nothing to prove, so suppose $H$ is nonscalar. As derivations $[\cdot, A]$ with elements $A\in\mc{A}^n_1$ act as first order partial derivatives, there exists $A\in\mc{A}^n_1$ such that 
$$\mathrm{deg}([H,A])=\mathrm{deg}(H)-1.$$
But, $[H,A]=[Ud\pi(Z)U^*,A]\in\mc{A}^n_{k-1}$ so that $\mathrm{deg}(H)-1\leq k-1$, or $\mathrm{deg}(H)\leq k$.
Thus, $Ud\pi(Z)U^*=H+\lambda1\in\mc{A}^n_{k}$.
\end{proof}

\begin{example}\label{e:kerr} The Kerr Hamiltonian $H^2$, where $H=\frac{1}{2}(Q^2+P^2)$, is a self-adjoint degree-4 polynomial in the canonical operators arising in nonlinear optics. One might expect its unitary evolution $U_t=e^{itH^2}$ to be a one-parameter family of gates in $\mathcal{C}_4^1$. However, this is not the case: the Schwartz topology on $\mc{S}(\R)$ is equivalently generated by the semi-norms $p_k(\psi)=\norm{H^k\psi}, \ \ k\in\mathbb{N}_0$ (see, e.g., \cite{RS1}), so each $U_t$ leaves $\mc{S}(\R)$ invariant (as a function of $H$). Then, by Proposition \ref{p:smooth}, $U_t\in\mc{C}_k^1\Rightarrow U_t\in\mc{C}^{\infty,1}_k$. However, the following (standard) computation shows $U_t(aP+bQ)U_t^*$ is not a polynomial for arbitrary $a,b,t\in\R$.

For $n\in\N_0$, let $|n\ra$ denote the Hermite-Gauss number state. We write $H=N+\frac{1}{2}$ where $N$ is the number operator, and $aP+bQ=\frac{1}{\sqrt{2}}((b-ia)A+(b+ia)A^*)$ where $A,A^*$ are the annihilation and creation operators, respectively. We compute
\begin{align*}
    U_tAU_t^*|n\ra&=e^{-it(n^2+n+\frac{1}{4})}U_tA|n\ra=e^{-it(n^2+n+\frac{1}{4})}\sqrt{n}U_t|n-1\ra\\
    &=e^{it(-n^2-n-\frac{1}{4})}e^{it(n^2-n+\frac{1}{4})}\sqrt{n}|n-1\ra\\
    &=e^{-2itn}\sqrt{n}|n-1\ra\\
    &=Ae^{-2itN}|n\ra.\\
\end{align*}
Similarly, $U_tA^*U_t^*|n\ra=e^{2itN}A^*|n\ra$.
In terms of $Q$ and $P$, on $\mathrm{span}\{|n\rangle\}$ we have
$$U(aP+bQ)U^*=\frac{1}{2}((b-ia)(Q+iP)e^{-it(Q^2+P^2-1)}+(b+ia)e^{it(Q^2+P^2-1)}(Q-iP))$$
which, due to the presence of the exponentials, is not a polynomial in $Q$ and $P$. Thus, there are $t\in\R$ for which $e^{itH^2}\notin\mc{C}^1_k$ for any $k\in\N$. Nevertheless, the above calculations suggest that, roughly speaking, $e^{itH^2}$ belongs to $\mc{C}^1_4$ \textit{to first order} (by taking a linear approximation of the exponential series $e^{it(Q^2+P^2-1)}$ and ignoring potential self-adjointness issues). Note that the same conclusion ($e^{itK}\notin\mc{C}^1_k$) applies to the normal ordered Kerr Hamiltonian $K=(A^*)^2A^2$ as $e^{itK}$ differs from $e^{itH^2}$ by a Clifford gate.
\end{example}

\begin{remark}\label{r:approx} Exact gate teleportation in continuous variable systems requires infinite-energy resources, so any realistic implementation will necessarily be approximate. This fact, together with the previous example motivates the study of an \textit{approximate} CV Clifford hierarchy.
\end{remark}

Another notable consequence of Proposition \ref{p:smooth} is that any $U\in\mc{C}^{\infty,n}_k$ induces (by conjugation) an automorphism of the Weyl algebra $\mc{A}^n$. While the structure of automorphisms of the general $n$-mode Weyl algebra is not fully understood, when $n=1$, the automorphism group admits an amalgamated free product structure \cite{FM,ML}. In the remainder of this section we utilize this group structure to show that all single mode gates in the smooth Clifford hierarchy are semi-Clifford:

\begin{theorem}\label{t:smoothSC} For all $k\in\N$, $\mc{C}^{\infty,1}_k=\mc{SC}^1_k$.
\end{theorem} 

We begin our presentation with an overview of the known group structure of $\mathrm{Aut}(\mc{A}^1)$. In \cite{Dix68}, Dixmier showed that $\mathrm{Aut}(\mc{A}^1)$ is generated by the subgroups $A$ of affine symplectic transformations ($\alpha((a,b)(P,Q))=S(a,b)(P,Q)+c1$, $S\in Sp(2,\C)$, $c\in\C$) and $E$ of elementary triangular automorphisms ($\alpha(Q)=aQ+c1, \alpha(P)=a^{-1}P+p(Q)$, for some $a,c\in\C$ and polynomial $p$). Moreover, $\mathrm{Aut}(\mc{A}^1)$ is canonically isomorphic to the subgroup $H$ of the group $G$ of polynomial automorphisms of the plane $\C^2$  whose Jacobian determinant is 1 \cite{ML}. The group $G$ is known to be generated by the groups $A_0$ of (unnormalized) affine transformations and $E_0$ of (unnormalized) elementary triangular automorphisms ($\alpha(Q)=aQ+c1, \alpha(P)=bP+p(Q)$, $a,b,c\in\C$ and polynomial $p$) \cite{J}. Moreover, $G=A_0\ast_{A_0\cap E_0}E_0$, the amalgamated free product of $A_0$ with $E_0$ over their intersection $A_0\cap E_0$ \cite{FM}. By the normalized determinant condition, one easily sees that
$A\cap (A_0\cap E_0)=A\cap E=E\cap(A_0\cap E_0).$
Consequently, the subgroup $H=A\ast_{A\cap E} E\cong\mathrm{Aut}(\mc{A}^1)$. 

The group $H$ admits a natural multiplicative involution $\sigma$ given by $\sigma(\alpha)(X)=\alpha(X^*)^*$, where the adjoint $X^*$ denotes the restriction of the Hilbert space adjoint to $\mc{S}(\R)$. Note that on the subgroups $A$ and $E$, $\sigma$ acts by conjugating the complex coefficients implementing the automorphisms. In what follows we will be concerned with the fixed point subgroup $H^\sigma$, which corresponds to real-coefficient polynomial automorphisms.

\begin{lemma}\label{l:Hsigma}$H^\sigma=A^\sigma\ast_{A^\sigma\cap E^\sigma}E^\sigma$.
\end{lemma} 

\begin{proof} The intersection $A\cap E$ is the group of lower triangular affine automorphisms $s=(T,c)$, where $T\in Sp(2,\C)$ is lower triangular and $c\in\C$ is the affine shift. For any $S=\begin{bmatrix}a & b\\c & d\end{bmatrix}\in Sp(2,\C)$ with $b\neq 0$, there is a lower triangular $T=\begin{bmatrix}ad/b-c& 0\\a & b\end{bmatrix}\in Sp(2,\C)$ and $R=\begin{bmatrix}0 & 1\\-1 & b/d\end{bmatrix}\in Sp(2,\C)$ for which $R^{-1}S=T$. Thus, any automorphism $a=(S,c)\in A$ belongs to $r(A\cap E)$ for some $r\in X_A:=\{\id\}\cup\bigg\{\begin{bmatrix}0 & 1\\-1 & b\end{bmatrix}\mid b\in\C^{\times}\bigg\}$. Note that $X_A$ is (globally) invariant under $\sigma$.

Similarly, any $e\in E$ belongs to $\alpha_p(A\cap E)$, where $\alpha_p(Q)=Q$ and $\alpha_p(P)=P+p(Q)$ for a polynomial of degree $\geq 2$ with zero linear and constant factors (which are absorbed into the intersection $A\cap E$). Let $X_E$ denote the set of such $\alpha_p$, which is again (globally) invariant under $\sigma$.

Now, writing $h\in H^\sigma$ as a reduced word
$$h=x_1x_2\cdots x_n s,$$
with each $x_i$ alternating between $X_A$ and $X_E$ and $s\in A\cap E$, we have
$$h=\sigma(x_1)\sigma(x_2)\cdots\sigma(x_n)\sigma(s).$$
By global invariance of $X_A$ and $X_E$ under $\sigma$ and uniqueness of reduced word decompositions, we must have $\sigma(x_i)=x_i$ for all $i$, and $\sigma(s)=s$. The amalgamated free product structure then follows.
\end{proof}

Let $\mc{T}$ be the Bass-Serre tree associated to the amalgamated free product $H^\sigma=A^\sigma\ast_{I}E^\sigma$, where $I=A^\sigma\cap E^\sigma$ \cite{Bass,Serre}. The vertices of the bipartite tree $\mc{T}$ are the cosets of $A^\sigma$ and $E^\sigma$ and the edges are the cosets of $I$. Specifically, edge $hI$ connects the vertices $hA^\sigma$ and $hE^\sigma$. Viewing each edge as an isometric copy of the interval $[0,1]$, the tree $\mc{T}$ becomes a complete metric space under the path metric. The group $H^\sigma$ acts isometrically on $\mc{T}$ in the obvious manner, and the geometry of the action is linked to word length in the group: if $h=x_1\cdots x_n s$ is the reduced word decomposition of $h$ (as in the proof of Lemma \ref{l:Hsigma}) and $m$ is the midpoint of the edge $I$ (connecting the roots $A^\sigma$ and $E^\sigma$) then the lengths between the consecutive steps $s\cdot m=m$, $x_ns\cdot m=x_n\cdot m$, $x_{n-1}x_ns\cdot m$, etc. are all 1, so that $d_{\mc{T}}(h\cdot m,m)=n=\ell(h)$, the word length of $h\in H^\sigma$.  
 
\begin{lemma}\label{l:Bass-Serre} Let $(\alpha_t)_{t\in\R}$ be a one-parameter group in $H^\sigma$ for which $\ell(\alpha_t)\leq 3$ for all $t\in\R$. Then $\alpha(\R)$ is conjugate by an element of length at most 1 to a subgroup of either $A^\sigma$ or $E^\sigma$.
\end{lemma}

\begin{proof} The bounded length assumption implies that the $\mathrm{diameter}(\alpha(\R)m)\leq 3$. Since $\alpha(\R)$ is a group of isometries on the complete CAT(0) space $\mc{T}$ with a bounded orbit, the Bruhat-Tits fixed point theorem \cite[Corollary II.2.8]{BH} implies that the (unique) center $c$ of $\alpha(\R)m$ is fixed. If $c$ lies in the interior of an edge, then $\alpha(\R)$ must globally fix the edge (by nature of the action), and since it preserves the type of vertices (whether $A^\sigma$ or $E^\sigma$) it must fix the entire edge (vertices included). Let $v$ be the vertex closer to $m$. Then $d_{\mc{T}}(v,m)\leq 3/2$, (as $d_{\mc{T}}(c,m)\leq 3/2$). This vertex $v$ is necessarily of the form $aE^\sigma$, for some $a\in A^\sigma$, or $eA^\sigma$ for some $e\in E^\sigma$. In any case, there exists $g\in E^\sigma\cup A^\sigma$ such that $g\alpha(\R)g^{-1}$ lies in $A^\sigma$ or $E^\sigma$. 
\end{proof}

We are now in position to prove the main result of the section, which utilizes the previous Lemma to constrain the structure one-parameter subgroups of $\mc{SC}_k^1$ (which arise naturally from the inductive hypothesis). This constrained structure together with the canonical commutation relations allows one to derive an explicit semi-Clifford decomposition of the pertinent unitary, albeit with lengthy computations.

\begin{proof}[Proof of Theorem \ref{t:smoothSC}] We proceed by induction. The result is clearly valid for $k=1,2$, and by Theorem \ref{t:SC}, even for $k=3$. 

Suppose $\mc{C}^{\infty,1}_k=\mc{SC}^1_k$, $k>3$ and let $U\in\mc{C}^{\infty,1}_{k+1}$. Then $U_t:=UT_tU^*$ is a one-parameter group in $\mc{SC}^1_k=\mc{C}^{\infty,1}_k$. By Proposition \ref{p:smooth}, $U_t\mc{A}_1^1 U_t^*\subseteq \mc{A}^1_{k-1}$, and consequently, $U_t\mc{A}^1 U_t^*\subseteq\mc{A}^1$ for all $t\in\R$. Thus, $\alpha(t)=\mathrm{Ad}(U_t)$ is a smooth one-parameter group of automorphisms of $\mc{A}^1$ which are necessarily fixed by the involution $\sigma$. Moreover, since each $U_t$ is semi-Clifford, $\ell(\alpha_t)\leq 3$ for all $t\in\R$. By Lemma \ref{l:Bass-Serre}, $\alpha(\R)$ is conjugate to either (i) $A^\sigma$ or (ii) $E^\sigma$ by an element $g\in H^\sigma$ of length (at most) 1.

\underline{Case (i)}: Let $g\in E^\sigma\cup A^\sigma$ be such that $g\alpha(\R)g^{-1}=A^\sigma$. Without loss of generality, we may assume $g\in E^\sigma\backslash A^\sigma$ and for every $t\in\R$, there is $a(t)\in A^\sigma$ such that $\alpha_t=g^{-1}a(t)g$. Suppose there is some $t\in\R$ for which $a(t)\notin I$. Then $g^{-1}a(t)g$ is a reduced word decomposition of $\alpha_t$. Since $\alpha_t$ is conjugation by a semi-Clifford gate, it can always be represented as $a_1e_0a_2$ for some $a_1,a_2\in A^\sigma$ and $e_0\in E^\sigma$ (by Proposition \ref{p:decomp}). If each $a_1,e_0,a_2$ are outside of $I$, then uniqueness of reduced word decompositions contradicts the fact that $g\notin A^\sigma$. If at least one of $a_1,e_0,a_2$ belong to $I$, then the length of $\alpha_t$ is at most 2, contradicting the length of $g^{-1}a(t)g$. Thus, $a(t)\in I$ for all $t\in\R$, and $\alpha(t)$ is a one-parameter subgroup of $E^\sigma$, so it suffices to consider:

\underline{Case (ii)}: Let $g\in E^\sigma\cup A^\sigma$ be such that $g\alpha(\R)g^{-1}\subseteq E^\sigma$. Then, without loss of generality we may assume $g\in A^\sigma\backslash E^\sigma$ and for each $t\in\R$ there is $e(t)\in E^\sigma$ such that $\alpha(t)=g^{-1}e(t)g$. 

Let $S_1\in Sp(2,\R)$ and $c_1\in\R$ be such that $g(aP+bQ)=S_1(a,b)(P,Q)+c_11$, $a,b\in\R$, and write 
$$e(t)(Q)=a_tQ + b_t1, \ \ \ e(t)(P)=a_t^{-1}P+p_t(Q)$$
for some polynomial $p_t$ of degree $\leq k-1$, $a_t\in\R\backslash\{0\}$ and $b_t\in\R$. Since $e(s+t)=e(s)\circ e(t)$, for all $s,t\in\R$, it follows that $a_{s+t}=a_sa_t$, $p_{0}(Q)=0=b_0$. 

Let $T_{b_t}$ be the translation operator of $b_t$, $S_t=\mathrm{diag}(a_t^{-1},a_t)\in Sp(2,\R)$ and $r_t(x)=a_{t}\int p_t(x) \ dx$ be an anti-derivative to $a_tp_t(x)$ (which has degree $\leq k$). Then 
$$\mu(S_t)((a,b)(P,Q))\mu(S_t)^*=S_t(a,b)(P,Q)=a_t^{-1}aP+a_tbQ, \ \ \ a,b\in\R,$$ 
so that
\begin{align*}e^{ir_t(Q)}\mu(S_t)T_{b_t}PT_{b_t}^*\mu(S_t)^*e^{-ir_t(Q)}&=e^{ir_t(Q)}a_t^{-1}Pe^{-ir_t(Q)}=a_t^{-1}\sum_{l=0}^{k-1}\mathrm{ad}_{r_t(Q)}^l(P)\\
&=a^{-1}_t(P+[r_t(Q),P])\\
&=a_t^{-1}P+p_t(Q)\\
&=e(t)(P).
\end{align*}
Also, 
$$e^{ir_t(Q)}\mu(S_t)T_{b_t}QT_{b_t}^*\mu(S_t)^*e^{-itr_t(Q)}=e^{ir_t(Q)}(a_tQ+b_t1)e^{-ir_t(Q)}=(a_tQ+b_t1)=e(t)(Q).$$
Thus, $\mathrm{Ad}(UT_tU^*)=\alpha(t)=\mathrm{Ad}(C_1^*e^{ir_t(Q)}\mu(S_t)T_{b_t}C_1)\in\mathrm{Aut}(\mc{A}^1)$, where $C_1\in\mc{C}^1_2$ satisfies $g=\mathrm{Ad}(C_1)$. Also, by smoothness of $e(t)$, it follows that each family $(r_t(Q))\in\mc{L}(\mc{S}(\R))$, $(a_t)\in\R$, and $(b_t)\in\R$ is smooth in their respective spaces, in which case $e^{ir_t(Q)}$, $\mu(S_t)$ and $T_{b_t}$ are smooth families of unitaries (on $\mc{S}(\R)$) for which $1=e^{ir_0(Q)}=\mu(S_0)=T_{b_0}$. Therefore $UT_tU^*=C_1^*e^{ir_t(Q)}\mu(S_t)T_{b_t}C_1$ (up to a smooth $t$-dependent phase, which we can absorb into the definition of $r_t(Q)$) is a smooth one-parameter group with self-adjoint generator
$$UPU^*=\frac{1}{i}\frac{d}{dt}(C_1^*e^{ir_t(Q)}\mu(S_t)T_{b_t}C_1)\bigg|_{t=0} \ \ \ \textnormal{on $\mc{S}(\R)$.}$$
Before we calculate the derivative, note that, by (\ref{e:quadratic})
$$\frac{1}{i}\frac{d}{dt}\mu(S_t)\bigg|_{t=0}=\frac{1}{i}d\mu\bigg(\begin{bmatrix}-a'(0) & 0\\0 & a'(0)\end{bmatrix}\bigg)=\frac{a'(0)}{2}(PQ+QP)$$
and 
$$\frac{1}{i}\frac{d}{dt}T_{b_t}\bigg|_{t=0}=\frac{d}{dt}e^{ib_tP}\bigg|_{t=0}=b'(0)P.$$
Then,
$$R_1(Q)=\frac{1}{i}\frac{d}{dt}r_t(Q)\bigg|_{t=0}$$
is a polynomial of degree $k$ such that
\begin{align*}UPU^*&=\frac{1}{i}\frac{d}{dt}(C_1^*e^{ir_t(Q)}\mu(S_t)T_{b_t}C_1)\bigg|_{t=0}\\
&=\frac{1}{i}C_1^*\bigg(\frac{d}{dt}r_t(Q)\bigg|_{t=0}+\frac{d}{dt}\mu(S_t)\bigg|_{t=0}+\frac{d}{dt}T_{b_t}\bigg|_{t=0}\bigg)C_1\\
&=C_1^*(R_1(Q)+a_1(PQ+QP)+b_1P)C_1,
\end{align*}
where we have let $a_1=a'(0)/2$ and $b_1=b'(0)$.

Now, applying the exact same argument to the smooth one-parameter group $UM_tU^*$ in  $\mc{SC}^1_k=\mc{C}^{\infty,1}_k$ allows us to write
$$UQU^*=C_2^*(R_2(Q)+a_2(PQ+QP)+b_2P)C_2$$
for some $C_2\in\mathcal{C}^1_2$, polynomial $R_2(Q)$ of degree $k$, and $a_2,b_2\in\R$.

Letting $V=C_2U$, we have
\begin{align*}
VPV^*&=C(R_1(Q)+a_1(PQ+QP)+b_1P)C^*\\
VQV^*&=R_2(Q)+a_2(PQ+QP)+b_2P,
\end{align*}
where $C=C_2C_1\in\mc{C}^1_2$. Since $i1=[VQV^*,VPV^*]$, when expanding the commutator, all degree 1 or higher terms must vanish. In particular, since $k\geq 3$, the highest degree term in $$[R_2(Q),CR_1(Q)C^*]=[R_2(Q),R_1(c_PP+c_QQ+c_11)]$$ must vanish, forcing $c_P=0$: there exist $j,l\geq 3$ and non-zero $c_j,d_l$ such that
\begin{align*}0&=[c_jQ^j,d_l(c_PP+c_QQ+c_11)^l]=c_jd_l\sum_{m=0}^{j-1}Q^m[Q,(c_PP+c_QQ+c_11)^l]Q^{j-1-m}\\
&=c_jd_lc_P\sum_{m=0}^{j-1}ilQ^m(c_PP+c_QQ+c_11)^{l-1}Q^{j-1-m}.
\end{align*}
Thus, $CR_1(Q)C^*=R_1(c_QQ+c_11)$ is again a polynomial of degree $k$ in $Q$. Expanding the linear action of $\mathrm{Ad}(C)$ on $PQ+QP$ and $P$, and absorbing the resulting powers of $Q$ into $CR_1(Q)C^*$ allows us to write
$$VPV^*=R_1(Q)+a_1P^2+b_1QP+c_1P+id_11$$
for some new coefficients $a_1,b_1,c_1,d_1\in\R$ and polynomial $R_1(Q)$ of degree $k$ (where we also used $PQ+QP=-i1+2QP$).

We now show that $a_1=a_2=b_1=0$. Expanding the commutator and freely using the relation $PQ+QP=-i1+2QP$ gives
\begin{align*}i1&=[VQV^*,VPV^*]=a_1[R_2(Q),P^2]+b_1[R_2(Q),QP]+c_1[R_2(Q),P]\\
&+2a_2[QP,R_1(Q)]+2a_2a_1[QP,P^2]+(2a_2c_1-b_1b_2)[QP,P]+b_2[P,R_1(Q)]\\
&=a_1(2iR_2'(Q)P+R_2''(Q))+ib_1QR_2'(Q)+ic_1R_2'(Q)-2ia_2QR_1'(Q)+4ia_1a_2P^2\\
&+i(2a_2c_1-b_1b_2)P-ib_2R_1'(Q).
\end{align*}
Equation coefficients of $P^2$, we must have $a_1a_2=0$. Equating coefficients of $P$, we have
\begin{equation}\label{e:Q}i(2a_1R_2'(Q)-2a_2c_1-b_1b_2)=0.\end{equation}
If $a_1\neq0$, then $a_2=0$ and $R_2'(Q)=\frac{b_1b_2}{2a_1}$ is constant, implying $\mathrm{deg}(R_2(Q))\leq 1$ (contradiction), so we must have $a_1=0$. In this case, $2a_2c_1=b_1b_2$.

Equating remaining coefficients,
\begin{equation}\label{e:coeff}1=b_1QR_2'(Q)+c_1R_2'(Q)-2a_2QR_1'(Q)-b_2R_1'(Q).
\end{equation}
Supposing $b_1\neq 0$, we have $b_2=\frac{2a_2c_1}{b_1}$, and we can factor (\ref{e:coeff}) into
$$1=(b_1Q+c_1)(R_2'(Q)-\frac{2a_2}{b_1}R_1'(Q))$$
By functional calculus, this operator equation implies
$$1=(b_1x+c_1)(R_2'(x)-\frac{2a_2}{b_1}R_1'(x)), \ \ \ x\in\R,$$
taking $q=-\frac{c_1}{b_1}$ leads to a contradiction. Thus, $b_1=0$, and (\ref{e:coeff}) becomes
$$1=c_1R_2'(Q)-2a_2QR_1'(Q)-b_2R_1'(Q).$$
Since $\mathrm{deg}(QR_1'(Q))\geq k\geq 3$, it follows that $a_2=0$.

Summarizing these simplifications, 
$$VPV^*=R_1(Q)+c_1P+id_11, \ \ \ VQV^*=R_2(Q)+b_2P, \ \ \ 1=c_1R_2'(Q)-b_2R_1'(Q).$$
Since all nontrivial operators in this first equation are symmetric, we must have $d_1=0$. Differentiating the third equation,
$$c_1R_2''(Q)-b_2R_1''(Q)=(c_1R_2(Q)-b_2R_1(Q))''=0.$$
Hence, $c_1R_2(Q)-b_2R_1(Q)=Q+b1$, for some $b\in\R$ (the $Q$-coefficient is necessarily 1). We cannot have both $c_1,b_2=0$, so assume (without loss of generality) that $c_1\neq0$. Then $$R_2(Q)=c_1^{-1}b_2R_1(Q)+c_1^{-1}Q+c_1^{-1}b1, \ \ \textnormal{and} \ \ S:=\begin{bmatrix}c_1 & b_2\\0 & c_1^{-1}\end{bmatrix}\in Sp(2,\R).$$
Defining $R(x)=c_1^{-1}\int R_1(x) \ dx$, we obtain a degree $(k+1)$-polynomial for which 
\begin{align*}VPV^*&=e^{iR(Q)}\mu(S)T_{c_1^{-1}b}(P)T_{c_1^{-1}b}^*\mu(S)^*e^{-iR(Q)}\\
VQV^*&=e^{iR(Q)}\mu(S)T_{c_1^{-1}b}(Q)T_{c_1^{-1}b}^*\mu(S)^*e^{-iR(Q)}
\end{align*}
It follows that $U=C_2^*V =\alpha C_2^*e^{iR(Q)}\mu(S)T_{c_1^{-1}b}\in\mc{SC}^1_{k+1}$, for some phase $\alpha\in\T$.
\end{proof}

\section*{Acknowledgements}

J. C. was partially supported by the NSERC Discovery Grant RGPIN-2023-05133.

\section*{AI statement}

The authors acknowledge use of Gemini 3.6 in the Summer of 2026 and ChatGPT 6.1 in the Fall of 2026 to provide references and to help refine some of the ideas and calculations in the proofs of Theorems \ref{t:SC}, \ref{t:classify} and \ref{t:smoothSC}. We take full responsibility for the contents of the paper.

\end{document}